\documentclass[fleqn,usenatbib]{mnras}

\usepackage{newtxtext,newtxmath}

\usepackage[T1]{fontenc}
\usepackage{xcolor}

\usepackage{soul}
\let\VANthebibliography\thebibliography
\def\thebibliography{\DeclareRobustCommand{\VAN}[3]{##3}\VANthebibliography}

\usepackage{graphicx}	
\usepackage{amsmath}	
\usepackage{siunitx}
\usepackage{comment}
\usepackage{float}
\usepackage{placeins}
\usepackage{pifont}
\usepackage{footnote}
\usepackage[normalem]{ulem}

\newcommand{\udm}[1]{~{\rm #1}}
\newcommand{\cmfast}{\texttt{21cmFAST}\,}
\newcommand{\cmsense}{\texttt{21cmSense}\,}

\newcommand{\fstar}{$f_{*}$ }

\title[Effects of FFBs on 21-cm signal and reionization]{Effects of feedback-free starburst galaxies on the 21-cm signal\\ and reionization history}

\author[Libanore Sarah et al.]{%
Sarah Libanore,$^{1}$\thanks{E-mail: libanore@bgu.ac.il}
 Jordan Flitter,$^{1}$
 Ely D. Kovetz,$^{1}$
 Zhaozhou Li,$^{2}$
 and Avishai Dekel$^{2,3}$
\\
$^{1}$ Department of Physics, Ben-Gurion University of the Negev, Be'er Sheva 84105, Israel\\
$^{2}$ Racah Institute of Physics, The Hebrew University, Jerusalem 91904 Israel\\
$^{3}$ SCIPP, University of California, Santa Cruz, CA 95064, USA
}
\date{Accepted XXX. Received YYY; in original form ZZZ}

\pubyear{2023}

\begin{document}

\include{preamble}

\label{firstpage}
\pagerange{\pageref{firstpage}--
}
\maketitle

\begin{abstract}
Different star-formation models at Cosmic Dawn produce detectable signatures in the observables of upcoming 21-cm experiments.
In this work, we consider the physical scenario of feedback-free starbursts (FFB), according to which the star-formation efficiency (SFE) is enhanced in sufficiently massive halos at early enough times, thus explaining the indication from the James Webb Space Telescope for an excess of bright galaxies at $z \geq 10$. 
We model the contribution of FFBs to popII SFE and compute the impact these have on the 21-cm global signal and power spectrum. We show that FFBs affect the evolution of the brightness temperature and the 21-cm power spectrum, but they only have a limited effect on the neutral hydrogen fraction. We investigate how the observables are affected by changes in the underlying star formation model and by contribution from popIII stars. Finally, we forecast the capability of next-generation Hydrogen Epoch of Reionization Array (HERA) to detect the existence of FFB galaxies via power spectrum measurements. Our results show the possibility of a significant detection, provided that popII stars are the main drivers of lowering the spin temperature.  Efficient popIII star formation will make the detection more challenging.
\end{abstract}


\section{Introduction}

Between recombination and reionization, the Universe was permeated with neutral hydrogen. The absorption of CMB photons, collisions between particles and radiation from the first stars excited some of the hydrogen atoms from the singlet to the triplet state. These processes, together with hyperfine transitions that brought the atoms back to the singlet state, sourced the cosmological 21-cm signal (see e.g.,~Refs.~\citep{Furlanetto:2006jb,Pritchard:2011xb} for review). Its redshift evolution can be used to probe the conditions of the gas in the intergalactic medium (IGM)  across cosmic time.

While at very high redshift, in the so called Dark Ages, the 21-cm signal provides direct access to fluctuations in the matter density field, at Cosmic Dawn below $z \sim 30$ it becomes particularly sensitive to astrophysical processes related to the formation of the first stars and galaxies. The way neutral HI gas heats and ionizes in this phase depends on the efficiency of star formation and it develops inhomogeneously~\citep{Mesinger:2010ne,Munoz:2021psm}. Different star formation scenarios may lead to very different reionization histories, which can potentially be probed by next generation 21-cm experiments targeting the global signal (e.g., EDGES~\citep{Bowman:2018yin} and SARAS~\citep{Patra:2014qta}) or interferometers measuring the power spectrum, such as HERA~\citep{DeBoer:2016tnn}, MeerKAT~\citep{Wang:2020lkn} and SKA~\citep{Ghara:2016dva}.

This perspective is particularly timely, as recent JWST observations seem to suggest an anomalous large number of bright galaxies at high $z$~\citep{Naidu_2022,Donnan_2023,Labbe_2023,Bouwens:2022gqg,Mason:2022tiy}. Possible explanations point either to inconsistencies in the cosmological $\Lambda$CDM model~\citep{Lovell:2022bhx,Sabti:2023xwo,Hirano:2023auh,Padmanabhan:2023esp,Parashari:2023cui} or to the presence of uncertainties in the astrophysical model. Solutions that have been suggested within the $\Lambda$CDM paradigm usually contain {\it ad-hoc} prescriptions to match the observations, e.g.,~providing enhanced star-formation efficiency~\citep{Haslbauer:2022vnq}, large luminosity-to-mass ratio due to top heavy intial mass functions~\citep{steinhardt2023templates,Zackrisson:2011ct,Inayoshi_2022} or high UV radiation~\citep{yajima2022}, low dust attenuation~\citep{Fiore_2023,Wang:2023xmm}, or stochasticity in the star-formation history~\citep{Pallottini:2023yqg}. 

At the center of our work, we consider instead the scenario of feedback-free starbursts (FFB), which was proposed by the authors of ~\citet{Dekel:2023ddd}. Under the conditions of high density and low metallicity expected at high redshifts in massive dark-matter (DM) halos, star formation is predicted to be enhanced by an increased efficiency in converting the accreted gas into stars. At other epochs and DM halo masses, stellar feedback ---namely supernovae, stellar winds, radiative pressure and photo-heating--- lead to lower efficiency. FFBs naturally emerge when the free-fall timescale for star formation is $\sim \mathcal{O}(1\,{\rm Myr})$, i.e.,~shorter than the time required for a starburst to generate effective stellar feedback. ~\citet{Li:2023xky} further developed the FFB model, showing how it affects different observables during Cosmic Dawn, and simulations are currently under development, aimed at refining the theoretical detail of this scenario.
While giving rise to a high abundance of bright galaxies at $z \gtrsim 10$, the FFB scenario may also leave imprints on the 21-cm signal and reionization process. 

The synergy with 21-cm surveys, therefore, can be the key to provide observable tests for the FFB scenario.  
The goal of this work is to investigate the effect of FFB galaxies in this context.
In section~\ref{sec:model} we summarize the modelling required: we introduce the observables (\ref{sec:21cmsignal}), and how they depend on star formation (\ref{sec:SFR}) in the standard and FFB scenarios. Section~\ref{sec:analysis} describes the HERA survey characteristics and the setup of our analysis. In section~\ref{sec:FFB_signatures} we investigate how FFBs affect the 21-cm observables, namely the global signal and power spectrum. We account for contributions from population III stars in section~\ref{sec:MCG}. Finally, we discuss how this analysis translates to constraints on the detectability of the FFB scenario, which we forecast for HERA 21-cm power spectrum, in section \ref{sec:fisher}.  We draw conclusions in section~\ref{sec:conclusions}.


\section{Model}\label{sec:model}

To perform our analysis, we need first of all to introduce the 21-cm observables and to characterize how they depend on the underlying cosmological and astrophysical models. Crucial in this sense are the role of star formation and its efficiency; therefore, we summarize the main features of its model in the standard and FFB scenarios.

\subsection{21-cm observables}\label{sec:21cmsignal}

The main observables used to analyze 21-cm surveys are the brightness temperature~\citep{Barkana:2000fd} 
\begin{equation}\label{eq:Tb}
    T_{b} = \frac{T_s-T_\gamma}{1+z}(1-e^{-\tau_{21}}),
\end{equation}
and its fluctuations $\delta T_b$. In the previous equation, $T_\gamma\propto (1+z)$ is the CMB temperature and 
\begin{equation}
    \tau_{21}= (1+\delta) \,x_{\rm HI}\,\frac{T_0}{T_s}\frac{H(z)}{H(z)+\partial_r v_r}(1+z),
\end{equation}
is the 21-cm optical depth, which depends on the matter fluctuations $\delta$, the fraction of neutral hydrogen $x_{\rm HI}$ and the comoving gradient of the baryon peculiar velocity along the line-of-sight $\partial_r v_r$. The dependence on the cosmological model is collected into the Hubble parameter $H(z)$ and the normalization factor
\begin{equation}
    T_0 = 34\udm{mK}\left(\frac{1+z}{16}\right)^{1/2}\left(\frac{\Omega_bh^2}{0.022}\right)\left(\frac{\Omega_mh^2}{0.14}\right)^{-1/2}.
\end{equation}
Cosmological parameters $\{h,\Omega_b,\Omega_m,A_s,n_s\}$ are set at the {\it Planck 2018}~\citep{Planck:2018nkj} fiducial values throughout this work.

The last ingredient in Eq.~\eqref{eq:Tb} is the spin temperature $T_s$, that quantifies the ratio between the number density of hydrogen atoms in the triplet and singlet states. At thermal equilibrium, the spin temperature is set by 
\begin{equation}\label{eq:Ts}
    T_s^{-1} = \frac{x_\gamma T_\gamma^{-1}+x_cT_k^{-1}+x_\alpha T_\alpha^{-1}}{x_\gamma + x_c + x_\alpha}\,,
\end{equation}
where $T_k$ is the gas kinetic temperature and $T_\alpha\sim T_k$~\citep{Field:1959} is the color temperature of the Ly$\alpha$ photons emitted by the surrounding stars. The coefficient $x_\gamma \simeq 1$ couples the spin temperature to the CMB, while $x_c,\,x_\alpha$ couple it to the gas temperature. Following~\citet{Mesinger:2010ne}, $x_c$ depends on particle collisions and it can be estimated as a function of the number densities of neutral hydrogen, free electrons and free protons; its effect is relevant in the IGM only at $z\gtrsim 30$~\citep{Loeb:2003ya}. 
On the other hand, $x_\alpha$ is set through the Wouthuysen-Field process~\citep{Wouthuysen:1952,Field:1958,Hirata:2005mz} to be proportional to $ J_\alpha(\mathbf{x},z)/(1+z)$, namely to the Ly$\alpha$ background flux due to the integrated star formation rate. The value of $J_\alpha$ depends on HI excitation due to X-rays~\citep{Pritchard:2006sq} and to resonant scatterings in the Ly$\alpha$ series~\citep{Barkana:2004vb}. Near the sources, the HI optical depth and the contribution of high energy photons redshifted into the Ly$\alpha$ band make this coupling highly efficient. 
On the other hand, X-rays heat the gas faster; their luminosity is parametrized through a power-law, with a low-energy cut-off below which photons are absorbed before reaching the IGM~\citep{Fragos:2013bfa}. 

The Ly$\alpha$ (UV) radiation produced by astrophysical processes also leads to HI ionization. Initially, the process is balanced by the recombination rate~\citep{Sobacchi:2014rua,Park:2018ljd}, which stalls the growth of the ionized regions. Once the number of ionizing photons becomes high enough to saturate the Ly$\alpha$ coupling and make the interstellar medium transparent to other ionizing photons, these escape into the IGM~\citep{Verhamme_2015} and the fraction of neutral hydrogen $x_{\rm HI}$ in Eq.~\eqref{eq:Tb} decreases, leading to a decay in the 21-cm signal. 

All these processes arise inhomogeneously in the IGM: the amplitude and size of local fluctuations determine the 21-cm power spectrum, which is defined as
\begin{equation}\label{eq:power_spectrum}
    \Delta_{\rm 21cm}^2 = \frac{k^3}{2\pi^2}\langle \delta T_{b}\delta T_{b}^*\rangle.
\end{equation}

During the epoch of star formation,  Ly$\alpha$ photons initially couple the spin temperature to the adiabatically-decreasing kinetic temperature. Only after this coupling saturates, Ly$\alpha$ photons can heat the gas: this happens earlier in small DM halos, therefore small scales in the power spectrum have larger amplitude in this stage.
Large-scale power rises later, but it quickly overcomes the small scales due to X-ray heating, whose efficiency is larger close to individual sources, which are apart one from another. Once X-ray radiation reaches the IGM, fluctuations in the power spectrum are determined by the DM density field and, as time passes, by the morphology of HI ionized regions. Since ionization initially occurs due to UV radiation inside small DM halos~\citep{Wood:1999hx}, small scale power decreases faster. Once reionization is complete, the 21-cm signal disappears.


\subsection{Star Formation} \label{sec:SFR}

Star formation modelling is the key to understanding the 21-cm signal evolution at $z\! \lesssim\! 30$. Following~\citet{Munoz:2021psm} (MUN21), we consider a standard scenario in which reionization is driven by atomic cooling galaxies (ACGs) hosting population II (popII) stars, in agreement with faint galaxy observations and the UV luminosity function~\citep{Park:2018ljd,Behroozi:2014tna,Yung:2021met,Fialkov:2012su}. 
We then summarize the feedback-free starburst scenario~\citep{Dekel:2023ddd} (DEK23) and describe how it alters the star formation rate (SFR) and efficiency (SFE).

In both scenarios, we adopt the formalism from Refs.~\citep{Dekel:2013uaa,Dekel:2023ddd}, that characterizes the SFR per halo as 
\begin{equation}\label{eq:sfr_dek}
    {\rm SFR}(z,M_h) = f_{\rm duty}\epsilon(z,M_h)\dot{M}_{\rm acc}(z,M_h),
\end{equation}
where $\dot{M}_{\rm acc}(z,M_h)$ is the mean baryonic accretion rate, $ \epsilon(z,M_h)$ is the star formation efficiency, and $f_{\rm duty}\! =\! \exp\left(-{M_{\rm turn}}/{M_h}\right)$ includes a turnover mass $M_{\rm turn}$ to suppress the SFR on the small mass end. We approximate $\dot{M}_{\rm acc}$ using the analytical prescription in DEK23, 
\begin{equation}
    \dot{M}_{\rm acc} = 65\, M_\odot {\rm yr}^{-1}\left(\frac{M_h}{10^{10.8}M_\odot}\right)^{1.14}\left(\frac{1+z}{10}\right)^{5/2}.
\end{equation}
The SFR in Eq.~\eqref{eq:sfr_dek} differs from the approximated SFR model in MUN21 and relies on more informed galaxy formation studies (e.g., Refs.~\citep{Dekel:2013uaa,Dekel:2023ddd,Mason:2015cna,Tacchella:2018qny,Mirocha:2020slz}); in Appendix~\ref{sec:SFR_models}, we discuss in detail the difference between the two and their impact on 21-cm observables. 

In our analysis, we weight the SFR by the halo mass function\footnote{We adopt the $z$-dependent halo mass function from~\citep{Watson:2012mt}.} $dn/dM_h$, and we marginalize over $M_h$ to get the SFR density  
\begin{equation}\label{eq:sfrd}
    {\rm SFRD}(z) = \int dM_h \frac{dn}{dM_h}{\rm SFR}(z,M_h).
\end{equation}
The SFRD is the main quantity that enters the computation of the Ly$\alpha$ background and X-ray heating in the 21-cm signal; the shape of the halo mass function implies that the contribution of the more massive halos is suppressed compared to the small mass ones. 

Moreover, the SFRD enters the computation of the number of ionizing photons. As~\citet{Park:2018ljd} describes in detail, to compute it we need to introduce the parameter
\begin{equation}\label{eq:fesc}
    \tilde{f}_{\rm esc} = {\rm min}\left[f_{\rm esc}\left(\frac{M_h}{10^{10}M_\odot}\right)^{\alpha_{\rm esc}},1\right]\,,
\end{equation}
that describes the fraction of Ly$\alpha$ ionizing photons capable of leaving the galaxies and ionizing the intergalactic medium; we use $f_{\rm esc}=10^{-1.35}$, $\alpha_{\rm esc} =-0.3$. The value of $\tilde{f}_{\rm esc}$ in the Epoch of Reionization is still uncertain; recent results from CEERS~\citep{Mascia:2023} seem to point to a mismatch between data and theoretical prescriptions.

It is interesting at this point to note that, while the ionizing fraction depends on $\tilde{f}_{\rm esc}$, the heating is unaffected by its value~\citep{Park:2018ljd}. This is due to the longer mean free path that both soft-UV and X-ray photons that heat the gas have with respect to UV photons that drive reionization~\citep{Mesinger:2012ys,McQuinn:2012bq}. In fact, the cross section for the absorption of ionizing photons with energy $\geq 13.6\,$eV is very high: whenever the HI column density is large, they get trapped inside galaxies and are not capable of reaching the IGM. The recombination of HI atoms inside the galaxies then produces a Ly$\alpha$ cascade~\citep{Santos:2003pc} that adds to the bulk of soft-UV photons. Because of their lower energy, these can be absorbed only if the energy matches one of the lines in the Lyman series; the cross section of this process is smaller and results in a longer mean free path, that allows them to reach the IGM. Photons with energy $> 10.2\,$eV are later on redshifted into the Ly$\alpha$ line, and interact with HI and diffuse in the IGM as a result of  scattering due to their absorption and re-emission~\citep{Santos:2003pc}.

The main consequence of the different mean free paths of ionizing- and heating- photons is their dependence on the distribution of HI column density regions~\citep{Das:2017fys}. While the former is described by Eq.~\eqref{eq:fesc}, which leads to a suppression of the ionization in the more massive halos, the latter depends on the formation efficiency of the sources that mainly produce the radiation field. In our analysis, we consider three main drivers: popII stars, formed in atomic cooling galaxies; FFBs; popIII stars formed in molecular cooling galaxies. We characterize popII and FFB efficiency in the next subsections, while popIII stars are investigated in Sec.~\ref{sec:MCG}. 


\begin{figure*}
    \centering
\includegraphics[width=2\columnwidth]{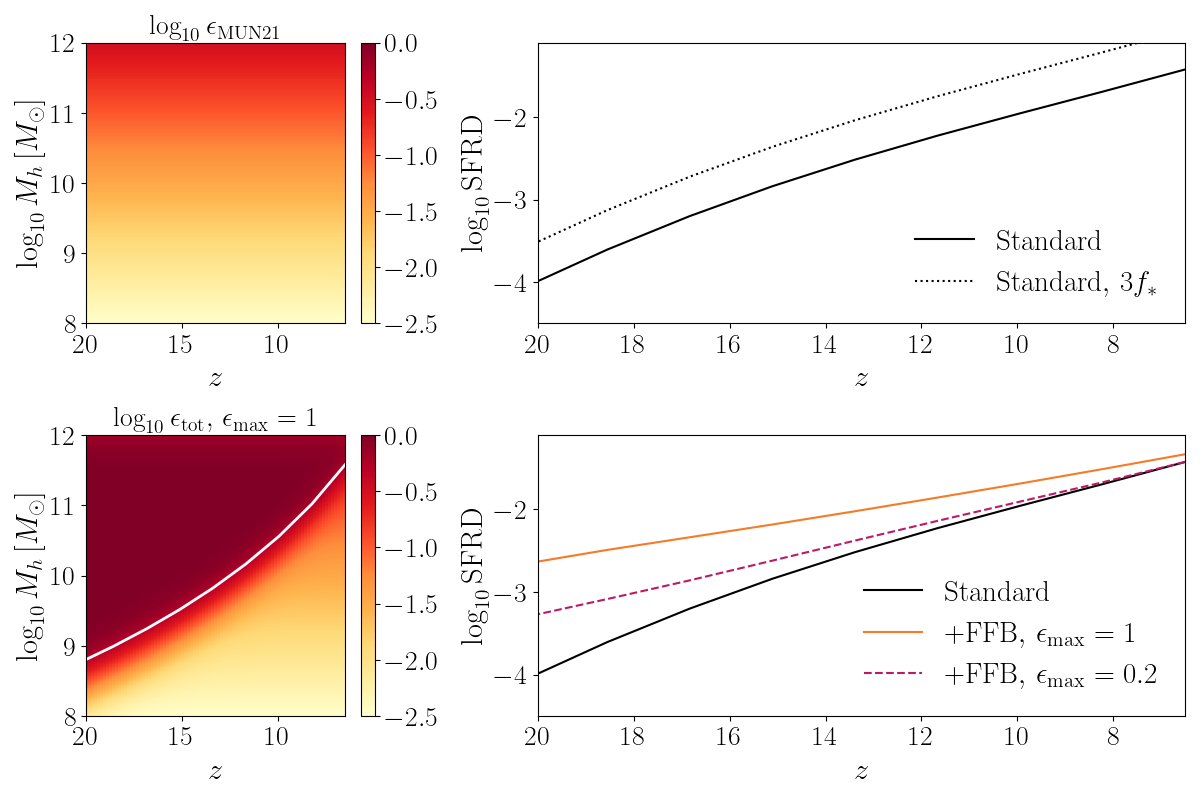}
    \caption{{\it Star formation models.} -- Left panels: SFE in the standard MUN21 case (top, Eq.~\eqref{eq:epsilon}) and when FFBs from DEK23 are included (bottom,  Eq.~\eqref{eq:eps_tot}) as a function of the halo mass and redshift. The white line indicates the threshold mass $M_{\rm FFB}$ for FFBs from Eq.~\eqref{eq:Mth}. Right panels: SFRD from Eq.~\eqref{eq:sfrd} marginalized over $M_h$ in the standard case using different $f_*$ values (top) and adding FFBs with different $\epsilon_{\rm max}$ (bottom). The fiducial value of \fstar in both the panels is $10^{-1.48}$. {\it FFBs affect star formation differently from just rescaling the efficiency.}}
    \label{fig:epsilon}
\end{figure*}

\subsubsection{Atomic Cooling Galaxies}\label{sec:ACG}

In the ACG scenario where popII stars are formed, we use the prescription in MUN21 to characterize the SFE in the standard case $\epsilon(z,M_h) = \epsilon_{\rm MUN21}(M_h)$, in which
\begin{equation}\label{eq:epsilon}
    \epsilon_{\rm MUN21}(M_h) = {\rm min}\left[f_*\left(\frac{M_h}{10^{10}M_\odot}\right)^{\alpha_*}, 1\right],
\end{equation}
where \fstar$= 10^{-1.48}$ 
sets the SFE in halos with pivot mass $M_h = 10^{10}M_\odot$.\footnote{We label as \fstar the factor that MUN21 calls $f_{*,10}^{\rm II}$, leaving the dependence on the mass scale $10^{10}M_\odot$ implicit. We note that our fiducial value for $f_*$ is lower than the one in MUN21, where \fstar$= 10^{-1.25}$ is used. We made this choice, as we show in Appendix~\ref{sec:SFR_models}, because the refined SFR model in Eq.~\eqref{eq:sfr_dek} leads to higher values than the approximated expression adopted in MUN21. The value \fstar$= 10^{-1.48}$ allows us to rescale the 21-cm signal to values estimated in previous works, while still being in agreement with current constraints~\cite{Abdurashidova_2022}.} The power law index $\alpha_* = 0.5$ is modelled as in~\citet{Wyithe:2012bq} to account for star-formation quenching in small DM halos. 

With respect to the SFRD in Eq.~\eqref{eq:sfrd}, we define $ M_{\rm turn} = {\rm max}(M_{\rm atom}, M_{\rm crit})$, where $M_{\rm atom}=3.3\times 10^{7}M_\odot [(1+z)/21]^{-3/2}$ is the minimum mass required to form stars via atomic cooling~\citep{Oh:2001ex}, while $M_{\rm crit}$ characterizes the critical halo mass below which star formation is inefficient because of photo-heating~\citep{Thoul:1996by,Noh:2014gra,Sobacchi:2014rua}. Fig.~\ref{fig:epsilon} shows the ACG SFE and SFRD we adopt in our analysis.


\subsubsection{Feedback-Free Starburst Galaxies}\label{sec:FFB}

When modelling star formation, the role played by stellar feedbacks, such as winds or supernova explosions, is crucial. The star formation efficiency at low $z$ is believed to be small due to feedback~\citep{Rodriguez_2017, Moster_2018,Behroozi:2012iw,Behroozi:2019kql}, while the FFB scenario introduced by DEK23 shows that the SFE is higher in massive galaxies at $z\sim 10$, in agreement with the excess of bright galaxies in JWST observations. 

For FFB to happen, the free-fall collapse time of the star-forming cloud (SFC) have to be shorter than the time required by the stellar feedbacks to become effective. The former is estimated as $t_{\rm ff} \propto n_{\rm SFC}^{-1/2}$, where $n_{\rm SFC}$ is the gas number density, while the latter is $t_{\rm fbk}\simeq 1\,{\rm Myr}$. 
Moreover, the timescale $t_{\rm ff}$ has to be larger than the time the gas requires to cool and form stars,~$t_{\rm cool}\propto n_{\rm SFC}^{-1}$~\citep{Krumholz:2012wi}. Finally, a large enough surface density $\Sigma_{\rm SFC} = {M_{\rm SFC}}/{\pi r_{\rm SFC}^2}$ is required to prevent unbounding the SFC gas through stellar radiative pressure and photo-ionization, $M_{\rm SFC}, r_{\rm SFC}$ being its mass and radius~\citep{Fall:2009nf,Grudic_2018,Grudic_2020}. 
In order for these processes to realize efficiently, not only do SFCs have to be free of their own feedbacks, but they also need to be shielded against UV radiation and winds from older-generation stars. 

The former is guaranteed, since $r_{\rm SFC}$ is larger than the ionizing length $\delta r$ inside which the UV photon flux overcomes the recombination rate.  
As for shielding against stellar winds (see e.g.,~\citet{Menon_2023}), the time a shock wave takes to cross the SFCs, namely the cloud crushing time~\citep{Klein:1994pd} $t_{\rm cc} \propto n_{\rm SFC}^{1/2}$, has to be longer than the timescales $t_{\rm ff},\,t_{\rm cool}$, previously introduced. Gas inside SFCs where FFB take place is almost completely consumed, so they reach near-zero column density~\citep{Menon_2023}. 

\noindent
It is possible that the evolution of the gas temperature, which is larger at higher $z$, has a non negligible impact on the onset of FFBs. However, its effect is not straightforward and it requires the use of numerical simulations to be understood, e.g.,~to model the way it alters the stellar mass function resulting from the fragmentation process. We defer to a future, dedicated work the study of this effect, while in this paper we rely on the simplified assumptions in DEK23.
All these conditions are satisfied only by short starbursts in DM halos continuously supplied by gas, all of which fragment into SFCs. 

\bigskip
As DEK23 shows, the criteria for the onset of FFB can be translated in terms of the properties of the host DM halo mass at given $z$.
Star formation in the halo is driven by FFB, thus its efficiency is enhanced with respect to the standard scenario when
\begin{equation}\label{eq:Mth}
    M_h \geq M_{\rm FFB}(z) = 10^{10.8}\left[(1+z)/{10}\right]^{-6.2}\,M_\odot.
\end{equation}
This threshold has been computed in DEK23 and it is shown in Fig.~\ref{fig:epsilon}

Star formation inside halos that satisfy the condition in Eq.~\eqref{eq:Mth} does not proceed with constant rate. As~\citet{Li:2023xky} describes in detail, during the FFB phase an halo undergoes multiple bursts of extremely high star formation, each of which consumes almost all the gas available in the star-forming clouds. The burst is hence followed by a period of quenched star formation, during which new gas accretes onto the galaxy, so to reach high enough density to fragment again. While each burst lasts a few Myr, namely a few times the free-fall timescale of the star-forming clouds, the interval between two subsequent bursts is $\sim 10\,{\rm Myr}$.  Overall, the conditions for FFBs can be materialized over a global period of $\sim 100$\,Myr, namely the time required to accrete $\sim 10^{10}\,M_\odot$ of gas, during which ten bursts can be realized, each leading to the formation of one generation of stars. DEK23 already showed that Eq.~\eqref{eq:Mth} ensures that each generation is shielded against feedback from the previous one. Following the approach in~\citet{Li:2023xky}, in the following we approximate the SFE during this time interval using a constant value, which averages the rate between the bursts and the times between them; this average quantity turns out to be larger than the SFR in the standard scenario.

Eq.~\eqref{eq:Mth} highlights the fact that halos of mass $\sim 10^{10.8}\,M_\odot$ at $z\sim 10$ can host FFBs; the threshold decreases at higher redshift, where their presence significantly affects star formation. 
At low $z$, instead, the threshold mass gets larger, but at the same time the onset of AGN feedback and the presence of hot circumgalactic medium quench star formation in halos $M_{h}\geq M_{\rm q} = 10^{12}M_\odot$. Thus, in the local Universe, FFBs are unlikely.
 
The way FFBs contribute to the total star formation rate per halo ${\rm SFR}_{\rm tot}(z,M_h)$, can be modelled as 
\begin{equation}\label{eq:ffb_sfr}
    {\rm SFR}_{\rm tot} = (1-f_{\rm FFB}){\rm SFR}_{\rm std} + f_{\rm FFB}{\rm SFR}_{\rm FFB},
\end{equation}
where ${\rm SFR}_{\rm FFB} = \epsilon_{\rm max}\dot{M}_{\rm acc}$, and the parameter $\epsilon_{\rm max}\leq 1$ describes the maximum SFE that FFB galaxies can reach. 
The $(z,M_h)$ dependence, which we left implicit for brevity, is encoded in 
\begin{equation}\label{eq:sigmoid}
\begin{aligned}
    f_\mathrm{FFB}(z,M_h) =\,\, & \mathcal{F}\times \mathcal{S}\left[\frac{\log_{10} {M_\mathrm{q}}/{M_{h}}}{0.15 \mathrm{dex}}\right] \times \\
    & \times \mathcal{S}\left[\frac{\log_{10}{M_{h}}/{M_\mathrm{FFB}(z)}}{0.15 \mathrm{dex}}\right]
\end{aligned}
\end{equation}
where $\mathcal{F} \leq 1$ is the fraction of galaxies that form in halos with $M_h > M_{\rm FFB}(z)$ and host FFBs, while $\mathcal{S}[x]=(1+\mathrm{e}^{-x})^{-1}$ is a sigmoid function varying smoothly from 0 to 1.
The first sigmoid characterizes the quenching for $M_h \geq  M_\mathrm{q}$, while the second sets the star formation rate to its value in the standard model ${\rm SFR}_{\rm std}$ for halos below the threshold $M_{\rm FFB}(z)$, while it gains a ${\rm SFR}_{\rm FFB}$ contribution in halos that host FFBs. 

The relation in Eq.~\eqref{eq:ffb_sfr} can be translated to a relation between the SFE in the standard case ($\epsilon_{\rm MUN21}$, from Eq.~\eqref{eq:epsilon}) and the SFE in galaxies where star formation is driven by FFBs. We consider\footnote{Eq.~\eqref{eq:eps_tot} is converted into Eq.~\eqref{eq:ffb_sfr} by using Eq.~\eqref{eq:sfr_dek} and assuming the same baryonic accretion rate for different scenarios $\dot{M}_{\rm acc}^{\rm tot}\!=\!\dot{M}_{\rm acc}^{\rm std}\!=\!\dot{M}_{\rm acc}^{\rm FFB} $.}
\begin{equation}\label{eq:eps_tot}
    f_*\epsilon_{\rm tot} = f_*(1-f_{\rm FFB})\left(\frac{M_h}{10^{10}M_\odot}\right)^{\alpha_*} + f_{\rm FFB}\epsilon_{\rm max},
\end{equation}
where we kept the $f_*$ normalization for consistency.
We set $\{\epsilon_{\rm max},\mathcal{F}\} = \{1,1\}$ as fiducial values, meaning that all the galaxies formed in halos with $M_h\geq M_{\rm FFB}(z)$ have SFE up to $\epsilon_{\rm max}= 1$. In the analysis, we keep $\mathcal{F} = 1$ fixed, but we test the more conservative case $\epsilon_{\rm max} = 0.2$, where FFB galaxies reach smaller efficiency.

The left panels of Fig.~\ref{fig:epsilon} compare the standard SFE $\epsilon_{\rm MUN21}(M_h)$ from Eq.~\eqref{eq:epsilon} with $\epsilon_{\rm tot}(z,M_h)$ from Eq.~\eqref{eq:eps_tot}, in which FFBs with $\epsilon_{\rm max} = 1$ are included. In the right panels, the figure compares the standard SFRD with the cases of interest for our analysis. 
Finally, in Appendix \ref{app:UV} we show how the high-$z$ UV luminosity function changes when FFBs are included. The figure can be compared with Fig. 5 in \citet{Li:2023xky}, where a more detailed discussion on this observable can be found.


\vspace*{-0.5cm}
\section{Analysis Setup}\label{sec:analysis}

The enhanced star formation efficiency from Eq.~\eqref{eq:eps_tot} naturally has a non-negligible impact on the 21-cm signal.

To estimate the effect of FFBs on 21-cm obervables, we customized the public code \cmfast\footnote{\url{https://github.com/21cmfast}, version 3.3.1, June 2023.}~\citep{Mesinger:2010ne,Munoz:2021psm,Murray:2020trn}.
The code simulates the reionization history by modelling the radiation fields we described in Sec.~\ref{sec:21cmsignal} and their effects on the thermal evolution and neutral hydrogen fraction inside cells of the simulation box. The evolution of cosmological fluctuations can be consistently accounted for via the initial conditions in \texttt{21cmFirstCLASS}\,\citep{flitterI,flitterII}.
We modified the ACG SFE by introducing the redshift-dependent $\epsilon_{\rm tot}$ from Eq.~\eqref{eq:eps_tot}; as described in the previous Section and in~\citet{Li:2023xky}, this quantity is used to approximate with a constant, average value the SFE during the $\sim 100\,{\rm Myr}$ in which an halo satisfies the condition set in Eq.~\eqref{eq:Mth} to host FFBs. We changed $\epsilon_{\rm tot}$ at each $z$ only for halos close to and above the mass threshold in Eq.~\eqref{eq:Mth}, according to Eq.~\eqref{eq:sigmoid}. Halos that exit the FFB condition behave as in the standard scenario, with no consequences due to their previous state. This is justified by the fact that, after $\sim 20\,{\rm Myr}$ from the last burst of star formation, most of the massive stars produced during the FFB phase have evolved and exited their active phase. Therefore, their presence does not increase the amount of feedback with respect to the standard case, and so SFE can be restored to its standard value.

Throughout the analysis, we used a (256\,Mpc)$^3$ simulation box, inside which 384 cells are defined on each axis for the high resolution computation, related with initial condition and displacement field, and 128 for the low resolution one, for temperature and ionization fluctuations~\citep{Mesinger:2007pd}. We used initial adiabatic fluctuations described by the approximation in~\citet{Munoz:2023kkg}, the \texttt{CLASS} configuration for the matter power spectrum, the $z$-dependent mass function from~\citet{Watson:2012mt}, and we included the effects of redshift space distortions and relative velocities between dark matter and baryons. The code smooths the density perturbations over neighbouring cells; finally, Ly$\alpha$ heating is also included~\citep{Sarkar:2022dvl} (while CMB heating is not). 

We estimate the global signal and neutral hydrogen fraction based on lightcones produced by \cmfast, and compute the 21-cm power spectrum with \texttt{powerbox}\footnote{\url{https://github.com/steven-murray/powerbox}}~\citep{Murray_2018}.


\subsection{Noise model}\label{sec:detector}

We compute the 21-cm power spectrum noise when observed with HERA, the Hydrogen Epoch of Reionization Array.\footnote{\url{http://reionization.org/}} HERA is a next-generation radio interferometer, a low-frequency precursor of SKA in South Africa. Its final configuration will comprise of 350 parabolic dishes, 14m diameter size each, arranged in a hexagonal configuration. It will observe frequencies between 50 to 250 MHz to probe the redshifted 21-cm emission from $z\!\sim\! [5,27]$~\citep{DeBoer:2016tnn}. Currently, Phase I HERA data have been released using $\sim 50$ antennas and 18 hours of observations to probe the power spectrum at $z = 7.9$ and $ 10.4$ and scales $k \sim 0.19\,h/{\rm Mpc}$ and $0.34\,h/{\rm Mpc}$~\citep{HERA:2021bsv,HERA:2022wmy,Lazare:2023jkg}.

Forecasted HERA sensitivity depends on the detector configuration and on the goodness of foreground removal. Following Refs.~\citep{Parsons:2011ew,Pober:2012zz,Pober:2013jna}, the baseline length between each pair of detectors sets the transverse modes $k_\perp$ that can be observed, while the bandwidth sets $k_\parallel$ along the line-of-sight. 
Spectral-smooth foregrounds (i.e.,~Galactic and extragalactic synchrotron and free-free emissions) mainly contaminate small $k_\parallel$. Their contribution, however, is then processed through the chromatic response function of the instrument: the mode-mixing sourced by the interferometer itself can deteriorate the signal on different angular locations as a function of frequency. It has been shown (see e.g.,~\citep{Datta:2010pk}), that this effect is mainly relevant for large $k_\perp$: this determines a wedge-shape in the $(k_\perp,k_\parallel)$ plane. As a result, the 
values of $k_\parallel$ associated with small $k_\perp$ are mostly foreground-free.~\citet{Pober:2013jna} defines different foreground removal models depending on the shape of the wedge edge: in this work, we will adopt the ``moderate" and ``optimistic" foreground removal scenarios. The former extends the edge to $0.1 \,h{\rm Mpc}^{-1}$ beyond the horizon
\begin{equation}
    k_{\parallel}^{\rm hor} = {2\pi}{|\vec{b}|}/({Yc}),
\end{equation}
where $|\vec{b}|$ is the baseline length, $c$ the speed of light and $Y = c(1+z)^2/\nu_{\rm 21}H(z)$ is the conversion factor between bandwidth and line-of-sight distance, while $\nu_{\rm 21} \sim 1420.4\,{\rm MHz}$ is the rest frame frequency associated with the 21-cm line. The optimistic model, instead, improves the constraining power on both the small and large scales by extending $k_\parallel$ to the FWHM of the primary beam, computed as ${\rm FWHM} = 1.06\lambda_{\rm obs}/14\,{\rm m}\sim 10^\circ$. Both  models assume to add coherently different baselines, namely the integration times are summed when the same pixel is sampled more than once by redundant baselines. More detail on the noise computation is given in Refs.~\citep{Pober:2012zz,Pober:2013jna}.

\begin{figure*}
    \centering
   \includegraphics[width=\columnwidth]{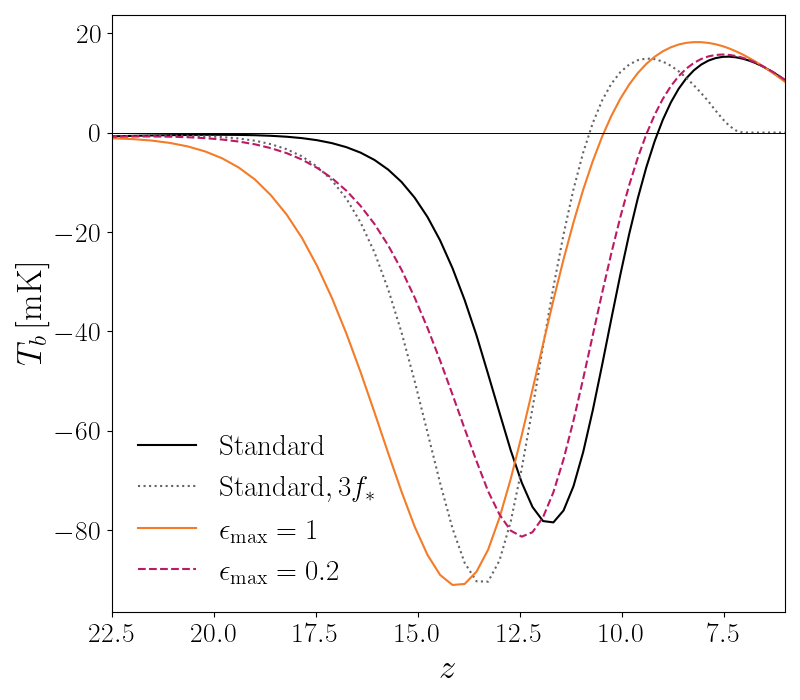}
   \includegraphics[width=\columnwidth]{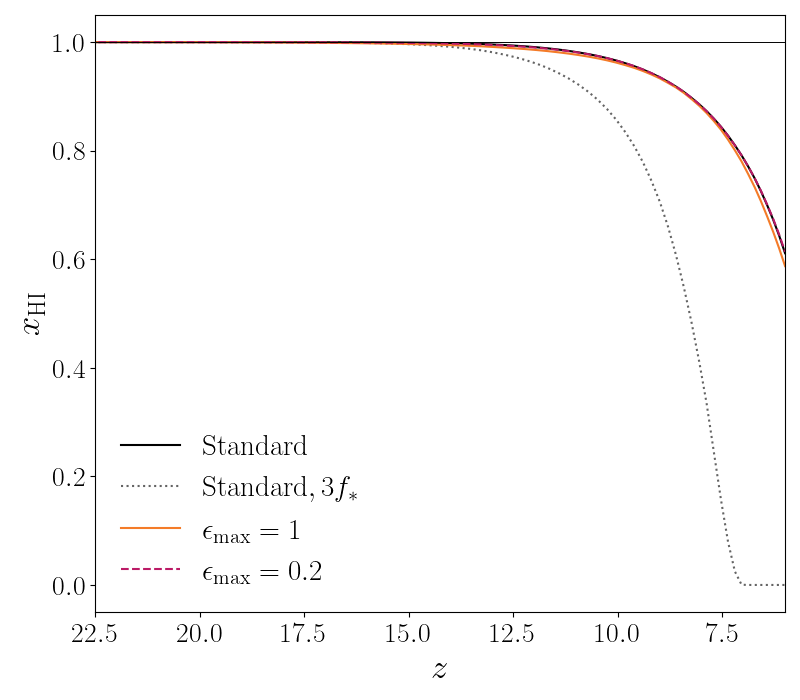}
    \caption{{\it FFB effect on $T_b$ and $x_{\rm HI}$.} -- $T_b$ global signal (left) and neutral hydrogen fraction $x_{\rm HI}$ (right) in the standard scenario using either the nominal \fstar (black) or $3$\fstar (gray, dotted), compared with the case that includes FFBs with $\epsilon_{\rm max}$ = 1 (orange) or $0.2$ (magenta, dashed). {\it FFBs anticipate the $T_b$ peak, while they have negligible effect on $x_{\rm HI}$, due to low $\tilde{f}_{\rm esc}$ in massive halos.} }
    \label{fig:GS_popII}
\end{figure*}

To estimate the HERA sensitivity, we rely on the public code \cmsense\footnote{\url{https://github.com/steven-murray/21cmSense.app}}~\citep{Pober:2012zz,Pober:2013jna}, which combines the contribution from the thermal noise power spectrum 
\begin{equation}\label{eq:thermal}
    \Delta^2_{\rm th} \sim \frac{k^3}{2\pi^2}\frac{X^2Y\Omega}{t}T_{\rm sys}^2,
\end{equation}
and the sample variance.
Here, $X$ converts the observed angles into transverse measurements, $T_{\rm sys}$ is the system temperature, $t$ the duration of the observational run and $\Omega = 1.13\, {\rm FWHM}^2$ the solid angle associated with the primary beam. The sample variance instead is estimated as the 21-cm power spectrum in Eq.~\eqref{eq:power_spectrum} and summed with the thermal noise to get the noise variance $\sigma_{\rm HERA}^{2}$.

In our analysis, we consider a hexagonal configuration with 11 dishes per side (i.e.,~331 antennas in total), each having 14m diameter. In Eq.~\eqref{eq:thermal}, the observational time $t$ is set to 6 hours per day over 540 days, while $T_{\rm sys} = T_{\rm sky}+T_{\rm rcv}$, where the sky temperature is $T_{\rm sky} = 60{\,\rm K}/(\nu/300\,{\rm MHz})^{2.55}$ and the receiver temperature is $T_{\rm rcv} = 100{\, \rm K}$. We consider a minimum observed frequency of 50\,MHz, a maximum frequency of 225\,MHz and 8\,MHz bandwidths probed by 82 channels each. This sets the observed redshift bins to
\begin{equation}
    [z_0,z_{1}] = \left[\frac{\nu_{\rm 21}}{50\,{\rm MHz}}-1,\frac{\nu_{\rm 21}}{(50+8)\,{\rm MHz}}-1\right] , ...
\end{equation}
The 19 bins obtained are equally spaced in frequency but not in redshift, providing a finer sampling at low $z$.


\section{FFB signatures on 21-cm observables}\label{sec:FFB_signatures}

The analysis presented in this work has been realized using the public codes: \cmfast, version 3.3.1 updated in June 2023; \texttt{powerbox}; \cmsense. 
We modified \cmfast to include FFB galaxies and to account for the SFR formalism defined in Eq.~\eqref{eq:sfr_dek}. As we discuss in detail in Appendix~\ref{sec:SFR_models}, this SFR model differs from the approximation used in \cmfast public release: in the standard scenario, it provides a slightly larger SFR, thus anticipating reionization with respect to, e.g.,~results in MUN21.

We checked that, in the range of scales probed by HERA, using a single realization of the power spectrum or averaging over 5, 10 or 15 \cmfast simulations provides variations smaller than the error bars. Therefore, to reduce the computational cost, plots and forecasts are realized using the same random seed.

\subsection{Global signal}

First of all, we investigate how FFBs affect the 21-cm global signal. This observable, in fact, can help us understanding in a more straightforward way the peculiarities the FFB scenario has compared with other cases.

Fig.~\ref{fig:GS_popII} shows how the presence of FFBs impacts the brightness temperature and reionization once different values of $\epsilon_{\rm max}$ are considered. 
The value of $T_b$ in Eq.~\eqref{eq:Tb} is estimated for our FFB prescription in Eq.~\eqref{eq:eps_tot} and compared with the standard \cmfast configuration from Eq.~\eqref{eq:epsilon}. For comparison, we also consider an artificial toy model in which $\epsilon_{\rm MUN21}$ is increased over all the halo masses and the entire redshift range, by simply rescaling the value of \fstar by $3$; the SFRD for the same model is also shown in Fig.~\ref{fig:epsilon}. This value was chosen to match the star formation efficiency required by JWST observations at $z\sim 9$ without FFBs (see e.g., Refs.~\citep{Labbe_2023,Boylan-Kolchin:2022kae}), but has no particular physical meaning. 

When star formation efficiency is increased at high redshift, a larger amount of Ly$\alpha$ and X-ray radiation is produced, speeding up the coupling of the spin temperature to the gas temperature and anticipating the moment in which this heats up. Therefore, in the top panel of Fig.~\ref{fig:GS_popII}, both the 3\fstar and FFB cases induce a larger $T_b$ global signal and move its peak towards large $z$. 
Moreover, the increased Ly$\alpha$ flux enhances the efficiency of the coupling between the spin temperature $T_s$ and the gas temperature $T_\alpha\sim T_k$ in Eq.~\eqref{eq:Ts}. Even if the gas at higher redshift is still hot, the stronger coupling drives $T_s$ farther from the CMB temperature $T_\gamma$, leading to a larger difference in the computation of the brightness temperature in Eq.~\eqref{eq:Tb}. For this reason, both in the 3\fstar and in the FFB models in Fig.~\ref{fig:GS_popII} the $T_b$ absorption peak becomes deeper when shifted at higher $z$.
However, as more time passes, only the more massive halos keep satisfying the conditions that allow the presence of FFB, while SFR in halos with masses between $10^8\,M_\odot$ and $10^{10}\,M_\odot$ becomes less efficient. 

\begin{figure*}
    \centering
   \includegraphics[width=\columnwidth]{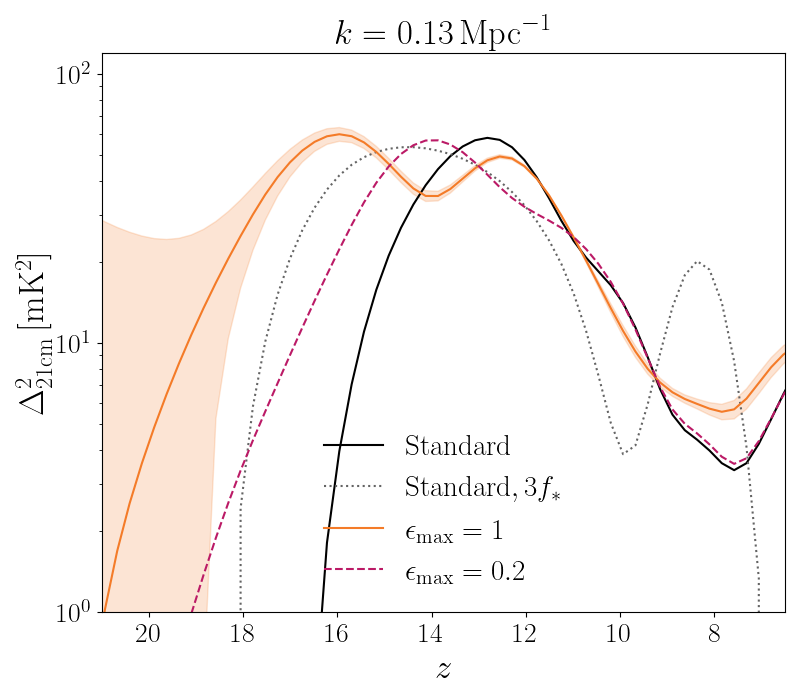}
   \includegraphics[width=\columnwidth]{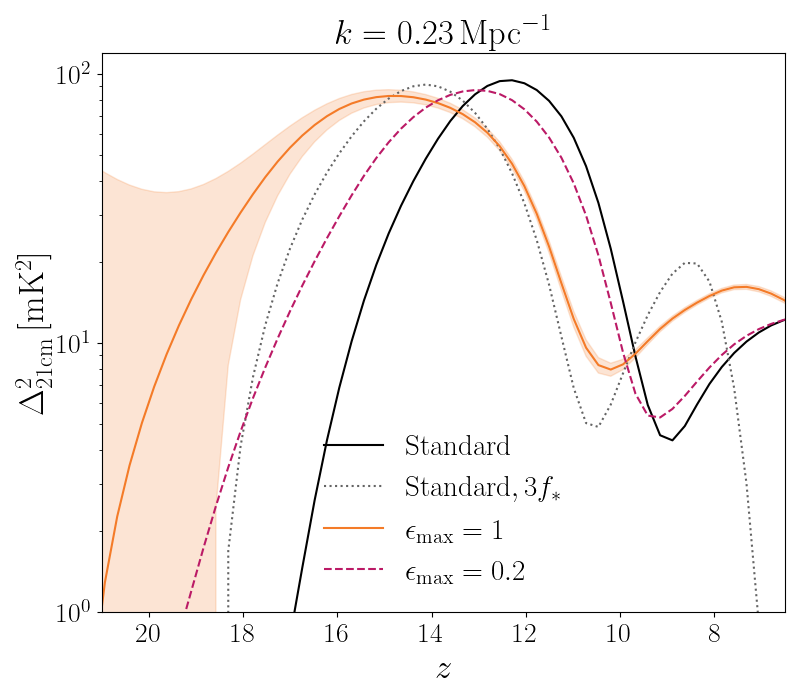}
    \caption{{\it FFB effect on $\Delta^2_{\rm 21cm}$.} -- Power spectrum as function of $z$; the scales we show are the ones HERA Phase I constrained~\citep{HERA:2021bsv,HERA:2022wmy}, once we set $h = 0.6736$ from {\it Planck 18}~\citep{Planck:2018nkj}. Same legend as Fig.~\ref{fig:GS_popII}. The shaded area shows $\pm \sigma_{\rm HERA}$, where the noise is computed for HERA with moderate foreground. {\it Below $z \simeq 17$, FFBs signatures can be detected outside the errorbars.}}
    \label{fig:pk_popII}
\end{figure*}

While one might expect  the high efficiency of the FFBs to result in a more efficient reionization, we find that it is not necessarily the case. As described in Sec.~\ref{sec:21cmsignal}, the reionization rate is determined by the escape  fraction $\tilde{f}_{\rm esc}$ of the ionizing photons. In the model underlying our assumption, the negative value of $\alpha_{\rm esc}$, motivated by Lyman-$\alpha$ forest and CMB data~\citep{Qin:2021gkn}, penalizes the contribution of large halos. Since these are indeed the one that host FFBs, the effect of FFBs on $x_{\rm HI}$ is limited. In contrast, if we were using positive values of $\alpha_{\rm esc}$ we would increase the ionizing power of massive halos and thereby improve the relevance of FFBs on reionization. 

In Appendix~\ref{app:lightcones} we show how the brightness temperature evolves inside lightcones produced by \texttt{21cmFAST}. The behaviour of the temperature in the lightcones and the presence of structures on the small scales reflect the evolution of the global signal and of the power spectrum, which we describe in the next Section. 

As a side comment, note that we adopted the same expression for the escape fraction $\tilde{f}_{\rm esc}$ in Eq.~\eqref{eq:fesc} for both non-FFB and FFB galaxies. This quantity is determined by the neutral hydrogen column density inside a galaxy, which describes the abundance of atoms capable of absorbing the ionizing radiation, preventing it to reach the IGM.
In principle, the presence of FFBs may increase $\tilde{f}_{\rm esc}$, since they consume all the gas in the star-forming clouds, and they remove dust via steady wind~\citep{Li:2023xky}. As the escape fraction in the Epoch of Reionization is anyway very uncertain, we leave further investigation on how it gets affected by FFBs to future work.

To sum up, when FFBs are taken into account, the peak in the 21-cm global signal starts at higher $z$, reaches a lower value and then ends slightly before the standard scenario; smaller values of $\epsilon_{\rm max}$ or $\mathcal{F}$ make the effect less significant in an almost-degenerate way. This is different from what we would expect for an overall increased SFR, as we model with 3$f_*$. Here, the peak shifts at higher $z$ but, thanks to the large efficiency in small mass halos, reionization is faster and the signal reaches $T_b = 0$ earlier.

\subsection{Power spectrum}

FFBs also affect the 21-cm power spectrum. 
Fig.~\ref{fig:pk_popII} shows its redshift evolution: consistently with
the global signal, the ionization bump rises earlier, at $z\sim 8$, for the 3\fstar model, while for FFBs it matches the standard scenario at $z\sim 6$. The presence of massive FFB-hosting halos increases $\Delta^2_{\rm 21cm}$ at high $z$; their lack of ionization power would keep the signal amplitude large even at low $z$, but the contribution of small halos brings $\Delta^2_{\rm 21cm}$ back to the standard case. The errorbas in the figure are estimated for HERA with moderate foreground, assuming the FFB scenario as fiducial, as in Sec.~\ref{sec:detector}; qualitatively, FFB signatures on the 21-cm power spectrum seems to be distinguishable in certain $(z,k)$ ranges.

\begin{figure*}
    \centering
   \includegraphics[width=2.\columnwidth]{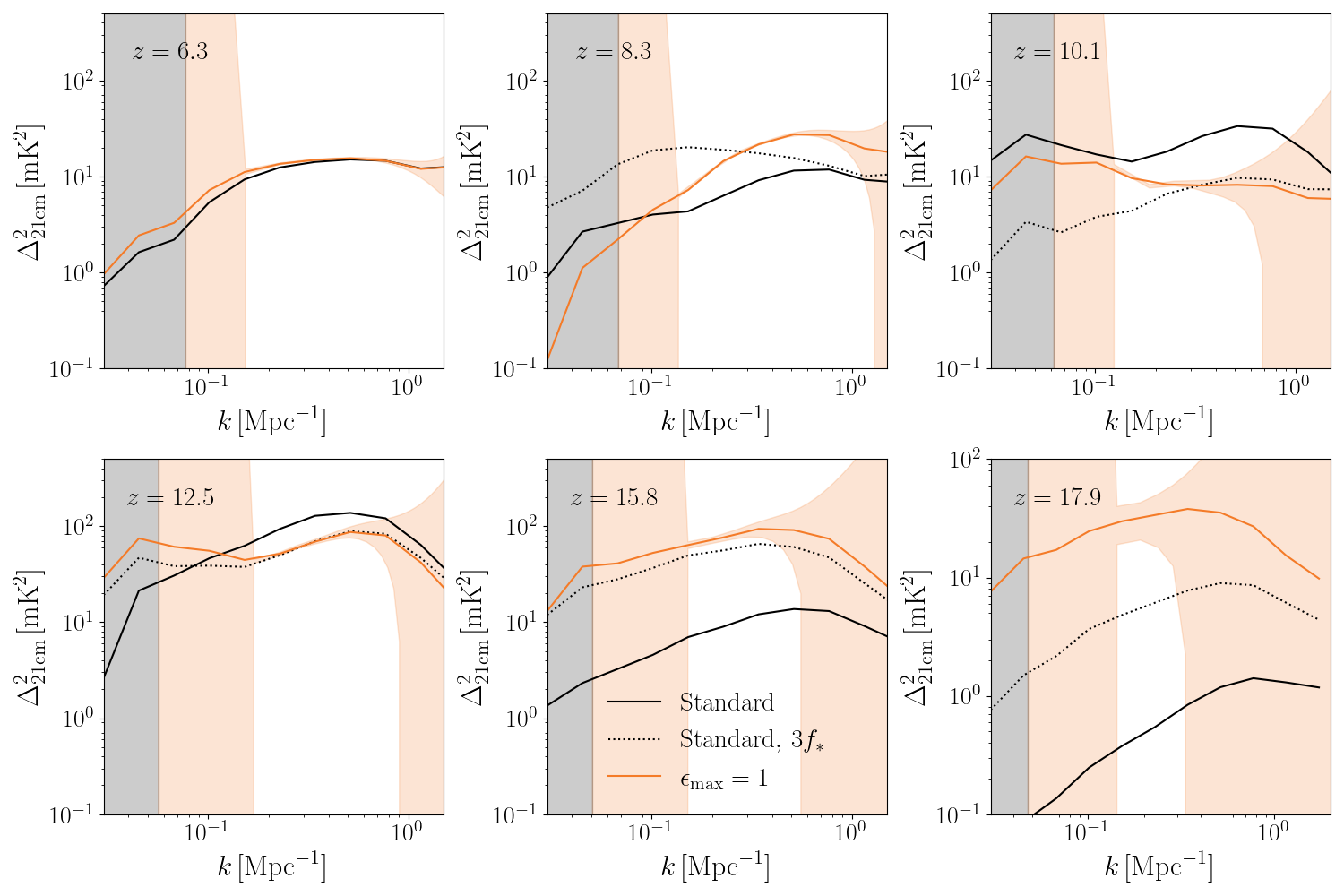}
    \caption{{\it FFB effect on $\Delta^2_{\rm 21cm}$ redshift evolution.} -- Power spectrum in the standard scenario (black continuous and dotted lines, respectively \fstar and 3\fstar) and including FFBs (orange, $\epsilon_{\rm max}=1$). The orange shaded area shows $\sigma_{\rm HERA}$, while the gray area indicates the $k-$range not probed by HERA. Here we run \cmfast in a larger, 700{\,Mpc} side box, to understand how FFBs contributes to larger scales. At $z=6.3$, there is no 21-cm power spectrum for $3f_*$ since reionization is complete. {\it FFBs boost large and then small scales at  $8<z<13$, since they form in the most massive and more clustered halos.}}
    \label{fig:pk_scales}
\end{figure*}

Finally, in Fig.~\ref{fig:pk_scales} we compare the redshift evolution of the scale-dependent power spectrum with and without FFBs. These plots were obtained using a 700\,Mpc simulation box to access larger scales. At very high redshift, the power spectrum in the FFB scenario has larger power; this can be understood comparing the amplitude of the global signal at the same epoch. At $8<z<13$, FFBs experience a boost initially on the large, then on the small scales: only the rare, most massive halos can still host FFBs at this redshift, thus increasing correlation on large scales; since these halos are the most densely clustered, they lead to an increase in the small scale power as well. Lower redshifts, instead, are dominated by the contribution of small halos; as already discussed, this brings the shape of the power spectrum back to the standard case.


\section{Including Molecular Cooling Galaxies}\label{sec:MCG}

A further contribution to the SFR and 21-cm signal could come from population III (popIII) stars~\citep{Bromm:2003vv,Bromm:2005gs}. 
Usually, popIII stars are associated with a pristine, metal-poor environment, and their formation is driven by H2 molecular cooling (see e.g.,~\citep{Haiman:1995jy,Haiman:1996rc,Abel:2001pr}). As in MUN21, we consider their formation as associated with molecular cooling galaxies (MCG) inside mini-halos, whose typical mass is $\sim 10^7M_\odot$. Their contribution to the reionization process is still uncertain, see e.g.~Refs.~\citep{Qin:2020xyh,Wise:2010zi,Xu:2016}, and not yet completely accepted. For example, recent results from HERA Phase I do not account for MCGs in their modelling; including them, depending on their efficiency, can lead to variations in the parameter constraints~\citep{Lazare:2023jkg}.

In this section, we model popIII contribution to SFE and we study how this affects the 21-cm observables, accounting for uncertainties. We assume MCGs cannot host FFBs, since the modelling in DEK23 refers to atomic cooling SFCs and the threshold mass in Eq.~\eqref{eq:Mth} penalizes mini-halos, its value being $M_{\rm FFB} > 10^7\,M_\odot$ up to $z \sim 40$. Thus, in our formalism, FFBs only enhance SFR for popII stars. The impact of different feedback levels in the formation of popIII stars, capable of accounting for all the uncertainties in the modeling of popIII star formation, requires a more detailed investigation. This however is beyond the scope of the current work; we thus limit our analysis by modeling differently popII (with and without FFB effects) and popIII, and dividing our results to the cases where popIII stars are either neglected or modeled in their standard scenario. We postpone to a follow-up paper the analysis of more complete cases, in which popIII stars can form in massive halos, or could be affected by FFBs. 


\vspace*{-.5cm}
\subsection{Model}\label{sec:MCG_modelling}

Following MUN21, we approximate SFR in MCGs as\footnote{See further discussion on this SFR approximation in Appendix~\ref{sec:SFR_models}.}
\begin{equation}\label{eq:sfr_popIII}
    {\rm SFR}^{\rm III}(z,M_h) = \frac{M_*^{\rm III}f_{\rm duty}^{\rm III}}{t_*H(z)^{-1}}=\frac{\epsilon^{\rm III}(z,M_h)f_bM_hf^{\rm III}_{\rm duty}}{t_*t_H(z)},
\end{equation}
where $t_* = 0.5$ is a fudge parameter and $t_H(z) = H(z)^{-1}$ is the Hubble time. The SFE is estimated as 
\begin{equation}\label{eq:eps_pop3}
    \epsilon^{\rm III}(M_h) = f_{*}^{\rm III}\left(\frac{M_h}{10^{7}M_\odot}\right)^{\alpha_{*}^{{\rm III}}}\,,
\end{equation}
where we set $\alpha_{*}^{{\rm III}}=0$ and we normalize with respect to the SFE in halos with $M_h=10^7M_\odot$, namely $f_{*}^{\rm III}$.\footnote{The parameter we indicate as $f_{*}^{\rm III}$ is called $f_{*,7}$ in MUN21.} 
\begin{figure*}
    \centering
   \includegraphics[width=\columnwidth]{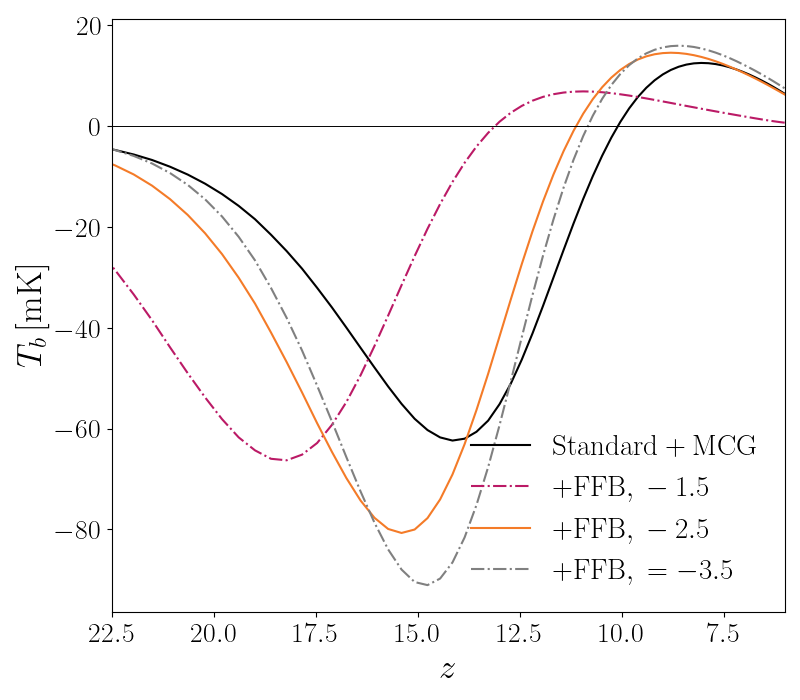}
   \includegraphics[width=\columnwidth]{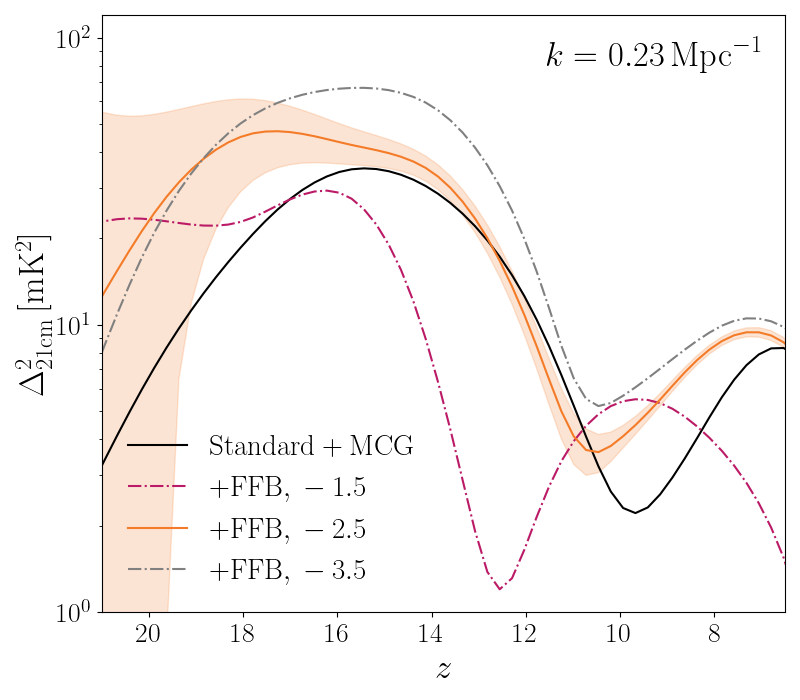}
    \caption{{\it PopIII contribution to $T_b$ and $x_{\rm HI}$.} -- Global signal (left) and power spectrum (right) when ACGs (popII), MCGs (popIII) and FFB contribute to SFR. We compare $f_{*}^{\rm III} = 10^{-2.5}$ to high $f_{*}^{\rm III} = 10^{-1.5}$ and low efficiency $f_{*}^{\rm III} = 10^{-3.5}$. {\it PopIII formation hides FFBs at high $z$, based on the efficiency of the process.}}
    \label{fig:GS_popIII}
\end{figure*}
We choose as nominal value $10^{-2.5}$; to account for uncertainties in the MCG efficiency, we also test $f_{*}^{\rm III} \in [10^{-1.5},10^{-3.5}]$. 
We use
\begin{equation}
\begin{aligned}
   &  f_{\rm duty}^{\rm III} = \exp\left(-\frac{M_{\rm turn}^{\rm III}}{M_h}-\frac{M_{ h}}{M_{\rm atom}}\right)\,,\\
        & M_{\rm turn}^{\rm III} = {\rm max}(M_{\rm mol}, M_{\rm crit}),
    \end{aligned}
\end{equation} 
to suppress star formation on the large halo mass end, where MCG star formation transits into the ACG scenario. The value of $M_{\rm mol} \propto f_{v_{\rm cb}}f_{\rm LW}$ accounts for quenching of star formation on the small mass end. In mini-halos, this is caused by the relative velocity between DM and baryons $v_{\rm CB}$~\citep{Tseliakhovich:2010bj,Naoz:2011if,Dalal:2010yt,Tseliakhovich:2010yw} and by Lyman-Werner feedbacks~\citep{Machacek:2000us,Kulkarni:2020ovu,Schauer:2020gvx} due to photons with energy between 11.2 and 13.6 eV, that photo-dissociate molecular hydrogen and prevent the cooling of the gas clouds. 
We fix the relative velocity contribution to 
\begin{equation}
    f_{v_{\rm cb}} = \left(1+A_{v_{\rm cb}}\frac{v_{\rm cb}}{v_{\rm rms}}\right)^{B_{v_{\rm cb}}},
\end{equation}
where $A_{v_{\rm cb}} = 1$, $B_{v_{\rm cb}} = 1.8$, the rms velocity is $v_{\rm rms} = v_{\rm avg}\sqrt{3\pi/8}$ and we set the average velocity to $v_{\rm avg} = 25.86\,{\rm km/s}$. 
As for LW feedbacks, we use~\citep{Visbal:2014fta}
$    f_{\rm LW} = 1 + A_{\rm LW}J_{21}^{B_{\rm LW}}$,
where  $A_{\rm LW} = 2$, $B_{\rm LW}= 0.6$ and $J_{21}$ is the LW intensity in units of $10^{-21}\,{\rm erg\,s^{-1}cm^{-2}Hz^{-1}sr^{-1}}$. 
The escape fraction from mini-halos is modelled analogously to Eq.~\eqref{eq:fesc}, with $f_{\rm esc}^{\rm III} = f_{\rm esc},\,\alpha_{\rm esc}^{\rm III} = \alpha_{\rm esc}$ and using $10^7M_\odot$ as the normalizing mass scale instead of $10^{10}M_\odot$.


\subsection{Effect on the 21-cm observables}\label{sec:MCG_signal}

The presence of MCGs inside mini-halos changes the 21-cm observables, mainly because of the larger radiation produced at high redshift. As Fig.~\ref{fig:GS_popIII} shows in the left panel, in the standard \cmfast case with MCGs, the global signal peak broadens and is preponed, leading also to an earlier reionization (although ACGs remain the main driver). Since popIII stars contribute also to the X-ray emission, their presence heats the gas faster, thus the $T_b$ peak becomes not as deep as in the only-ACG case. Analogously, the power spectrum shown in the right panel of Fig.~\ref{fig:GS_popIII} has larger power at high $z$, while it dies faster because of the earlier reionization.

The relevance of all these effects depends on the MCG star formation efficiency, encapsulated in the parameters $f_{*}^{\rm III}$ and $\alpha_*^{\rm III}$ in Eq.~\eqref{eq:eps_pop3}. In particular, following MUN21, we adopt $\alpha_*^{\rm III}=0$: this choice renders the popIII star formation efficiency independent from the mass of the mini-halos up to the turnover mass. If, instead, we had chosen $\alpha_*^{\rm III} < 0$, star formation in smaller halos would have been accelerated. On the other hand, effects due to large values of $f_*^{\rm III}$ partially cover the high-$z$ contribution of FFB galaxies in both the observables.
It is clear then that accounting for MCGs makes more challenging to detect the signatures of the FFB scenario.  


\vspace*{-0.2cm}
\section{Fisher forecasts}\label{sec:fisher}

In the previous sections, we estimated the effect of the existence of FFB galaxies on the 21-cm global signal and power spectrum. We now want to understand if HERA~\citep{DeBoer:2016tnn} will be able to detect the signatures of this scenario, provided the uncertainties on the MCGs contribution described in Sec.~\ref{sec:MCG}. For simplicity, we begin this analysis by ignoring the contribution of popIII stars; later, we relax this assumption in Sec.~\ref{sec:MCG_detectability}. Uncertainties related with the SFR model are discussed in Appendix~\ref{sec:SFR_models}.

To forecast the FFB detectability, we compute the Fisher matrix:
\begin{equation}\label{eq:fisher}
    F_{\alpha\beta} = \sum_{z,k} \frac{1}{\sigma_{\rm HERA}^2}\frac{\partial \Delta_{\rm 21cm}^2}{\partial\theta_\alpha}\frac{\partial \Delta_{\rm 21cm}^2}{\partial\theta_\beta},
\end{equation}
where derivatives are performed with respect to 
\begin{equation}
\theta = \{\epsilon_{\rm max},\log_{10}f_{*},\alpha_*,\log_{10}(L_X/{\rm SFR}),\log_{10}f_{\rm esc},\alpha_{\rm esc}\}.
\end{equation}
In the parameter set, $\epsilon_{\rm max}$ describes the properties of the FFB scenario, $\{\log_{10}f_*,\alpha_*\}$ characterize the ACG star-formation efficiency, $\{\log_{10}f_{\rm esc},\alpha_{\rm esc}\}$ the escape fraction and $\log_{10}(L_X/{\rm SFR})$ the X-ray luminosity. Degeneracies between the parameters are accounted via the process of marginalization; more details on their role in determining the 21-cm signal can be found in Sec.~\ref{sec:21cmsignal} and MUN21.
Fiducial values are summarized in Tab.~\ref{tab:fid}; we use uninformative priors on all the parameters. Variances $\sigma_{\rm HERA}^2$ are computed through \cmsense for the FFB scenario and including thermal noise and sample variance. The sum is performed over the $k$ bins computed by \cmsense and the 19 $z$-bins defined by HERA $8\,{\rm MHz}$ bandwidth.
\begin{table}
\renewcommand{\arraystretch}{1.5}
    \centering
    \begin{tabular}{|c|ccccc|}
    \hline
       $\epsilon_{\rm max}$ & $\log_{10}f_{*}$ & $\alpha_*$ & $\log_{10}(L_X/{\rm SFR})$ & $\log_{10}f_{\rm esc}$& $\alpha_{\rm esc}$  \\
    \hline
       $1; 0.2$ & {$-1.48$} & {$0.5$} & {$40.5$} & {$-1.35$} & {$-0.3$} \\
    \hline
    \end{tabular}
    \caption{Fiducial values in our Fisher forecast; for the FFB-related parameter $\epsilon_{\rm max}$ we consider two cases, as discussed in Sec.~\ref{sec:FFB}. Other cosmological and astrophysical parameters in \cmfast are fixed throughout the work. }
    \label{tab:fid}
\end{table}

\vspace*{-0.2cm}
\subsection{FFB detectability}\label{sec:forecasts}

\begin{figure}
    \centering
   \includegraphics[width=\columnwidth]{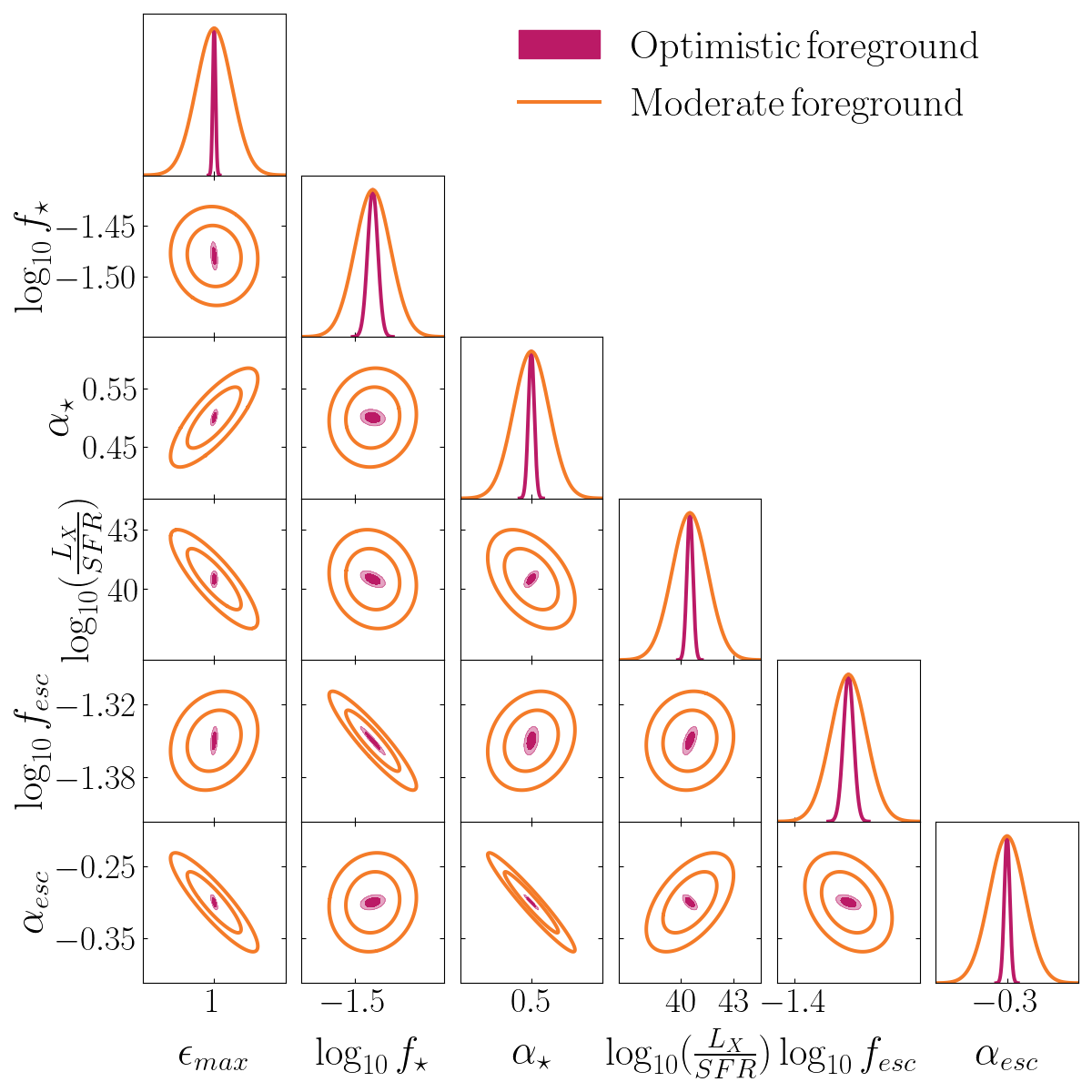}
    \caption{{\it Confidence ellipses.} -- Marginalized $1\sigma$ ellipses in the ACG+FFB case from Sec.~\ref{sec:forecasts} when $\epsilon_{\rm max}=1$. {\it The FFB parameter $\epsilon_{\rm max}$ has small degeneracies with other ACG-related parameters that affect the power spectrum. }}
    \label{fig:ellipse}
\end{figure}

First of all, we consider only the contribution of ACGs and FFBs, as described in Sec.~\ref{sec:model}. We estimate that, in the case of moderate foreground, the relative marginalized error on $\epsilon_{\rm max} = 1$ is $\sigma_{\epsilon_{\rm max}}/\epsilon_{\rm max} \simeq 6\%$; optimistic foreground improves the result to $\sigma_{\epsilon_{\rm max}}/\epsilon_{\rm max} \simeq 1\%$. Provided that the relative difference between $z = 10$ SFRD in the standard-ACG and FFB scenarios is $\sim \mathcal{O}(50\%)$, our analysis shows that the existence of FFBs can be detected with high significance both considering moderate and optimistic foreground removal. This was expected from the qualitative description in Sec.~\ref{sec:analysis}: FFBs have a relevant impact on the 21-cm power spectrum and unique features with respect to a simple enhancement of SFR. Thus, degeneracies between parameters entering the Fisher computation are tiny; we check that degeneracies are small between FFBs and other ACG-related parameters through the contour plot in Fig.~\ref{fig:ellipse}. 

Smaller values of $\epsilon_{\rm max}$ lead to closer SFRD in the two scenarios and to a weaker constraining power in the 21-cm analysis. For example, $\epsilon_{\rm max} = 0.2$ yields $\sigma_{\epsilon_{\rm max}}/\epsilon_{\rm max} \simeq 10\%$ with moderate foreground and $ \simeq 1\%$ with optimistic foreground, against a relative difference $\sim \mathcal{O}(10\%)$.\footnote{The absolute error is $\sim 0.02$ when $\epsilon_{\rm max}=0.2$, and $\sim 0.06$ when $\epsilon_{\rm max}=1$. The fact that the constraining power is comparable in the two cases can be explained by the non-trivial way the 21-cm power spectrum evolves as a function of $(z,k)$. For example, from Fig.~\ref{fig:pk_popII}, one can see that in some redshift ranges $\Delta_{21\rm cm}^2$ is larger for $\epsilon_{\rm max}=0.2$ than for $\epsilon_{\rm max}=1$. This leads to a larger signal in some of the observed bands, and to the possibility of getting good constrain when $\epsilon_{\rm max}=0.2$. Moreover, the Fisher matrix in Eq.~\eqref{eq:fisher} accounts for partial degeneracies with other astrophysical parameters; when $\epsilon_{\rm max}=0.2$, their effect can be constrained more easily, hence reducing the uncertainties and relatively improving the constraining power on $\epsilon_{\rm max}$. } We note that decreasing the fraction of FFB galaxies in the massive halos via the $\mathcal{F}$ parameter in Eq.~\eqref{eq:ffb_sfr} would lead to similar considerations, since its value is degenerate with $\epsilon_{\rm max}$.

This case represents our benchmark, providing the best results we can get assuming the SFR model is known and described by Eq.~\eqref{eq:sfr_dek}. Discussion on SFR model uncertainties can be found in Appendix~\ref{sec:SFR_models}.


\vspace*{-0.2cm}
\subsection{Effect due to MCG contributions}\label{sec:MCG_detectability}

We now account for contributions from popIII stars hosted by MCGs.
To do so, we compute the Fisher matrix including in the parameter set also $\{f_*^{\rm III},\alpha_*^{\rm III}\}$; other popIII-related parameters in \cmfast, namely $\{\log_{10}f_{\rm esc}^{\rm III},,\log_{10}(L_X/{\rm SFR})^{\rm III}\}$, are fixed to their fiducial values $\{{-1.35},40.5\}$ throughout the analysis.
While for the slope we consider as fiducial $\alpha_*^{\rm III}=0$, for the efficiency we test $\log_{10}f_{*}^{\rm III} \in [{-1.5},{-3.5}]$, 
to account for the large uncertainties on this parameter. The ``nominal" case assumes $\log_{10}f_{*}^{\rm III}$\,=\,$-2.5$; ``high efficiency MCGs" adopt $\log_{10}f_{*}^{\rm III}$\,$>$\,$ -2.5$; and finally ``low efficiency MCGs" consider $\log_{10}f_{*}^{\rm III}$\,$<$\,$-2.5$.  We discuss $\epsilon_{\rm max}$\,=\,$1$ for conciseness; smaller values lead to less stringent constraints, consistently with results discussed in Sec.~\ref{sec:forecasts}.

\begin{figure}
    \centering
    \includegraphics[width=\columnwidth]{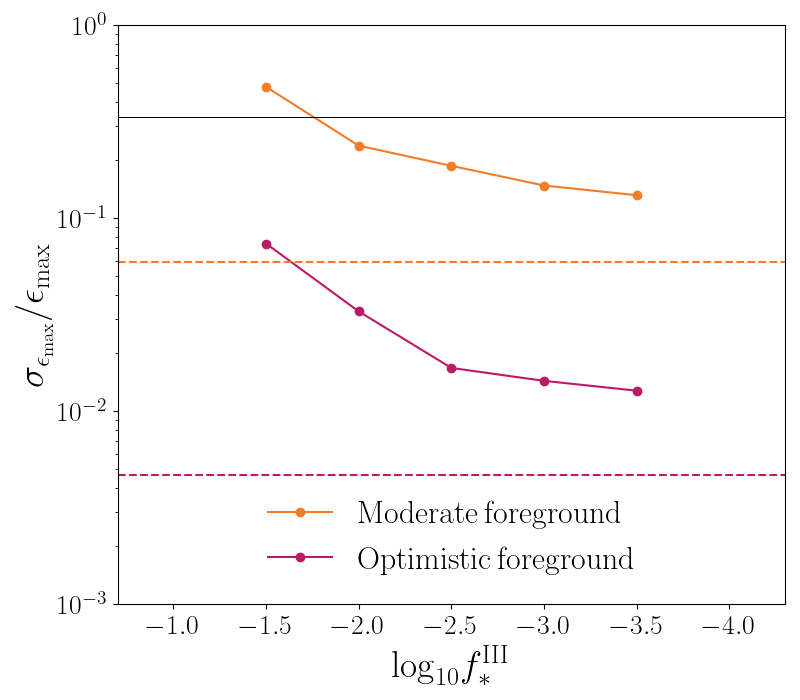}
    \caption{{\it Summary of our constraints on FFBs.} -- Marginalized $1\sigma$ error on $\epsilon_{\rm max}$ including MCG as a function of $\log_{10}f_{*}^{\rm III}$, with moderate (orange) and optimistic (magenta) foreground.  Horizontal dashed lines mark the case with only ACGs described in Sec.~\ref{sec:forecasts}. The black line shows $\sigma_{\epsilon_{\rm max}}/\epsilon_{\rm max} = 1/3$ as a reference, for which $\epsilon_{\rm max}$ can be detected $\sim\mathcal{O}(3\sigma)$. {\it FFB signatures can be detected by HERA when popIII star formation efficiency is not too high.}}
    \label{fig:sigma_A_MCG}
\end{figure}

Fig.~\ref{fig:sigma_A_MCG} collects our results on $\sigma_{\epsilon_{\rm max}}^{\rm III}$, i.e.,~the marginalized error on $\epsilon_{\rm max}$ once MCGs are included in the analysis.
In the case of moderate foreground, the presence of MCGs lowers the significance of the FFB detection: while with ``nominal" and ``low efficiency MCGs" values, FFB signatures can be still partially detectable, for ``high efficiency MCGs" the power spectrum becomes almost indistinguishable from the scenario without FFBs. The situation changes when optimistic foreground is considered: here, FFBs can be detected in both the ``nominal" and ``low efficiency MCGs" cases, while for ``high efficiency MCGs" the FFB detection is still plausible, even if with smaller significance. Even if the conditions for optimistic foreground are quite hard to reach, this result sets a benchmark for HERA's constraining power on FFBs. The two lines in Fig.~\ref{fig:sigma_A_MCG} hence pinpoint the reasonable detection level we will achieve with future full-HERA data-analysis, provided that foreground cleaning algorithms will reach the foreseen accuracy.  

\section{Conclusions}\label{sec:conclusions}

Upcoming years will provide improved measurements of the 21-cm global signal and power spectrum from the Epoch of Reionization. Combined with other probes, 21-cm experiments will shed light on the processes that regulate star formation in the first galaxies.
Uncertainties still exist on the star formation modelling, particularly regarding the role of popIII stars and stellar feedbacks, for which observations in the local Universe suggest an important role in quenching star formation efficiency. An extrapolation of feedback models to high redshifts should take into account other phenomena.

The authors of~\citet{Dekel:2023ddd} introduced the process of feedback-free starbursts, namely star formation events with short timescales that should arise in high redshift galaxies. For these to be efficient, gas clouds in which star formation takes place have to be dense enough and with low metallicity; these conditions guarantee that star formation has enough time to be realized before stellar feedbacks become effective. Moreover, under similar conditions, star-forming clouds would be shielded against radiation and winds from older stars. Overall, it is possible to show that these processes boost star formation efficiency inside halos above a certain mass threshold, whose value increases with cosmological time. Therefore, in the late Universe, feedback-free starbursts are rare since they can only be hosted by very massive halos; moreover, once AGN feedbacks set up, star formation always gets  quenched in halos $> 10^{12}M_\odot$. On the contrary, at high redshift the evolution of the threshold mass indicates that feedback-free starbursts can be found even in smaller halos; their presence could explain the existence of high redshift, massive galaxies observed by JWST. It is hence important to analyse further implications this model can have, so to understand which are the observable signatures that could either confirm it or rule it out.

Our work, together with~\citet{Li:2023xky}, represents one of the first steps in this direction. We 
investigated the observational signatures of the 
feedback-free starbursts 
on the 21-cm signal. We modelled their contribution to star formation efficiency in atomic cooling galaxies and implemented it into \cmfast to estimate their effect on the 21-cm global signal and 21-cm power spectrum. 

Our main results can be summarized as follows. 

\begin{itemize}
    \item The redshift and mass dependence of the SFE in the FFB scenario speed up the evolution of the brightness temperature and of the 21-cm power spectrum before $z\sim 15$. At lower redshift, instead, their evolution gets closer to the non-FFB scenario. These result respectively from the coupling between the spin and gas temperatures, and from the X-ray heating: the coupling is stronger at high $z$ when FFBs are accounted for, due to the low-mass halos that host FFB galaxies at those times; the heating, instead, gets effective at lower $z$, where only massive halos can still host FFBs. 
    
    \item On the other hand, the evolution of the neutral hydrogen fraction is only weakly affected by the presence of feedback-free starbursts. This is because the low-mass halos with high escape fraction of ionized photons host FFBs only prior to $z \sim 15$, practically before the onset of reionization. At lower redshift, such halos tend to be without FFBs, and they therefore contribute to reionization similarly to the standard scenario. On the other hand, the high-mass FFB galaxies at these later times have a negligible contribution to reionization because of their lower escape fraction.
    
\item We forecasted the detectability of the FFB scenario in the different regimes. We showed that future interferometers, such as HERA, will be able to detect signatures of their existence in the 21-cm power spectrum, compared with the standard scenario that only includes popII stars formed in atomic cooling galaxies. We also checked how our results change when the FFB efficiency is lower.

\item We accounted for the possible contribution at high redshift of popIII stars in molecular cooling galaxies and showed that this may hide the effect of FFBs. We drew forecasts as a function of popIII efficiency: our results show that, except for cases with high efficient popIII star formation, signatures of the FFB scenario can still be detected. The significance level will depend on the foreground level. 

\end{itemize}

To conclude, our work highlights the crucial role 21-cm experiments can have in testing astrophysical scenarios. Their synergy with other probes, such as JWST data, in the upcoming years will foster our research of the high redshift Universe, helping us to shed light on the puzzles related to reionization and the birth of the first galaxies.


\section*{Acknowledgments} 
\noindent SL~is supported by an Azrieli International Postdoctoral Fellowship. JF is supported by an ongoing Negev Scholarship by the Kreitman School at Ben-Gurion University.
EDK acknowledges support by Grant No.\ 2022743
from the US-Israel Bi-national Science Foundation (BSF) and Grant No.\ 2307354 from the U.S.\ National Science Foundation (NSF).
AD and ZL are supported by the Israel Science Foundation Grant ISF 861/20. 
ZL has received funding from the European Union’s Horizon 2020 research and innovation programme under the Marie Skłodowska-Curie grant agreement No 101109759 (``CuspCore'').
The authors thank Debanjan Sarkar, Lukas J.~Furtak and Brant Robertson for useful and stimulating discussions. We thank the anonymous referee for their comments, which helped us improving the clarity and quality of the manuscript.

\section*{Data Availability} 
Our analysis was performed using the publicly available codes \texttt{powerbox} (\url{github.com/steven-murray/powerbox}), \cmsense (\url{github.com/steven-murray/21cmSense.app}) and by customizing \cmfast (\url{github.com/21cmfast}, version 3.3.1, June 2023). Our version of the code, as well as the analytical modelling of the FFBs, are available upon reasonable request.

\bibliographystyle{mnras}
\bibliography{biblio.bib} 

\appendix

\section{Changing the star formation model}\label{sec:SFR_models}

Throughout the main text, all the ACG- and FFB-related results were obtained under the assumption that the SFR is computed based on Eq.~\eqref{eq:sfr_dek}, which from now on we label {\it nominal}. This is different from the prescription defined in MUN21 and defined in the \cmfast public release: as a first order approximation, these works assume that the SFR in the ACG scenario is
\begin{equation}\label{eq:sfr}
    {\rm SFR}_{\rm approx}(z,M_h) = \frac{M_*}{t_*H(z)^{-1}}=\frac{\epsilon(z,M_h)f_bM_h}{t_*H(z)^{-1}},
\end{equation}
with $M_*$ the stellar mass, $M_h$ the virial mass of the host halo, $\epsilon(z,M_h)$ defined in Eq.~\eqref{eq:epsilon} and $t_* = 0.5$. In MUN21 and related works, $\epsilon(z,M_h)$ is computed via Eq.~\eqref{eq:epsilon} assuming as fiducial values $f_* = 10^{-1.25},\,\alpha=0.5$; these values match the maximum posterior in the analysis of HERA Phase I results~\citep{Abdurashidova_2022}.

\begin{figure}
    \centering
   \includegraphics[width=\columnwidth]{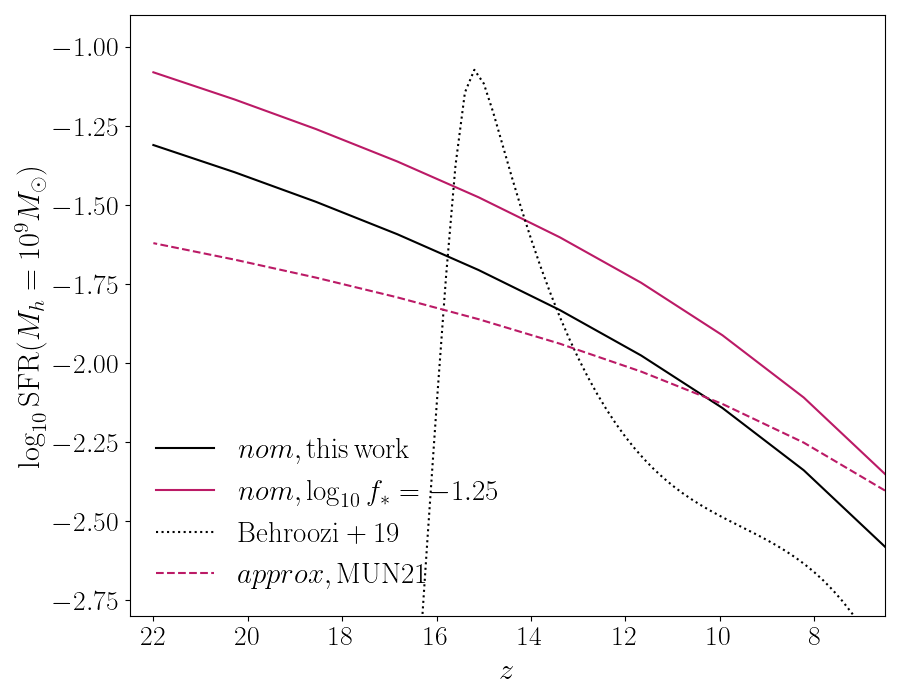}
    \caption{{\it Comparison between SFR formalisms.} --
    We compute the SFR in the standard scenario for halo with characteristic mass $M_h = 10^{9}M_\odot$. Solid lines indicate the {\it nominal} case, while dashed lines show the {\it approximated} scenario based on Eq.~\eqref{eq:sfr}. To compute the black line we used \fstar = $10^{-1.48}$ as in the main text, while for the magenta lines we adopted $10^{-1.25}$ as in MUN21. The dotted line refers to simulation results from~\citet{Behroozi:2019kql}. 
    {\it Different assumptions lead to different SFR redshift evolution.}}
    \label{fig:sfr}
\end{figure}

With respect to Eq.~\eqref{eq:sfr_dek}, this expression encodes a different SFR redshift evolution, as Fig.~\ref{fig:sfr} shows; in particular, the two SFR are comparable at low $z$, while the {\it nominal} case gets larger at higher $z$. Despite this, a similar SFRD can be recovered from Eq.~\eqref{eq:sfrd}; we showed this already in Fig.~\ref{fig:epsilon} in the main text. This is the most relevant quantity in the computation of the 21-cm signal.

In Fig.~\ref{fig:sfr}, we focus on the standard scenario\footnote{Including FFBs we would get similar results. We verified that the relative difference between the standard and FFB scenarios remains consistent when moving from the {\it nominal} SFR in Eq.~\eqref{eq:sfr_dek} to the {\it approximated} SFR in Eq.~\eqref{eq:sfr}.}  with $\epsilon(z,M_h)$ from Eq.~\eqref{eq:epsilon} and SFR computed using either Eq.~\eqref{eq:sfr_dek} or~\eqref{eq:sfr}. In order to get a similar SFR, the fiducial value of \fstar in the SFE computation needs to be changed between the two models, being lower when the SFR is computed via Eq.~\eqref{eq:sfr_dek}. This motivates our choice of using $f_* = 10^{-1.48}$ throughout the full analysis: this value remains compatible at $1\sigma$ with HERA Phase I results~\citep{Abdurashidova_2022} and it allows us to drive the SFR back to similar values to MUN21. The figure shows that, if the same \fstar than MUN21 was used, the {\it nominal} model would generally provide a larger SFR than the {\it approximate} one, thus leading to an earlier reionization. 

As we show in the next section, the choice of setting \fstar$=10^{-1.48}$ in the {\it nominal} case also allows us to recover a similar 21-cm global signal between the {\it nominal} and {\it approximated} SFR model. We verified that larger values of $\alpha$ could also help in matching the SFR results from the two prescriptions, boosting the efficiency for the rarest, more massive halos while decreasing it for the smaller ones. The importance of this parameter is anyway smaller with respect to \fstar, and, on the high mass side, it partially mimicks the effects of FFB we are interested in analysis; the degeneracy between these two parameters is caught inside the Fisher matrix analysis in Sec.~\ref{sec:analysis}. For these reasons, we preferred to set $\alpha$ to the same fiducial value adopted in MUN21, while matching the SFR by decreasing \fstar. 

In Fig.~\ref{fig:sfr}, we show as well the SFR obtained extrapolating at high-$z$ the results of~\citet{Behroozi:2019kql}, MUN21 commented on the different $\alpha$ dependence of its model from the one in~\citet{Behroozi:2019kql}, explaining the motivation as due to differences in the assumed star-formation histories. The prescription from~\citet{Behroozi:2019kql} becomes less reliable for $z\gtrsim 12$; since the high redshift range is crucial to model the 21-cm signal, we decided to avoid this approximation in our analysis in the main text.

\begin{figure*}
    \centering
   \includegraphics[width=\columnwidth]{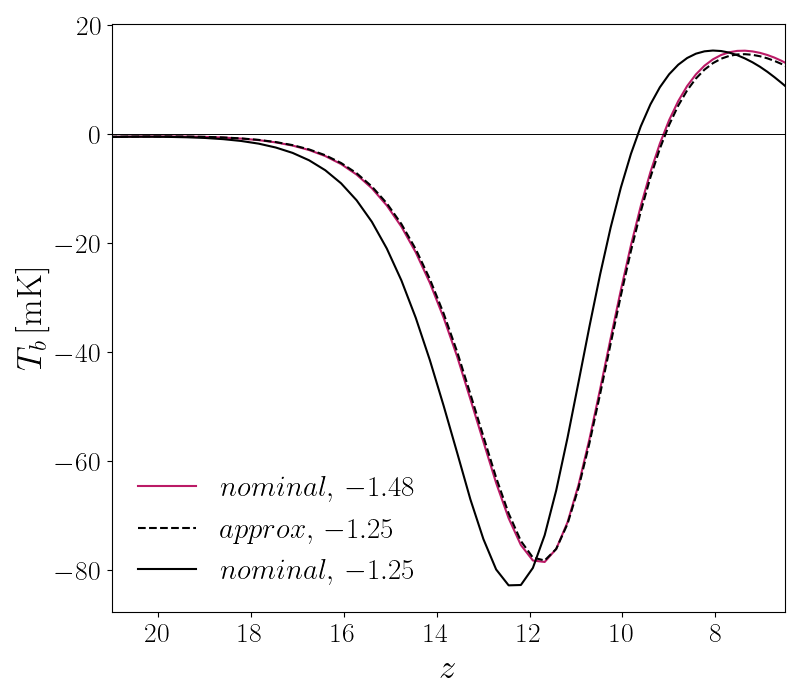}
   \includegraphics[width=\columnwidth]{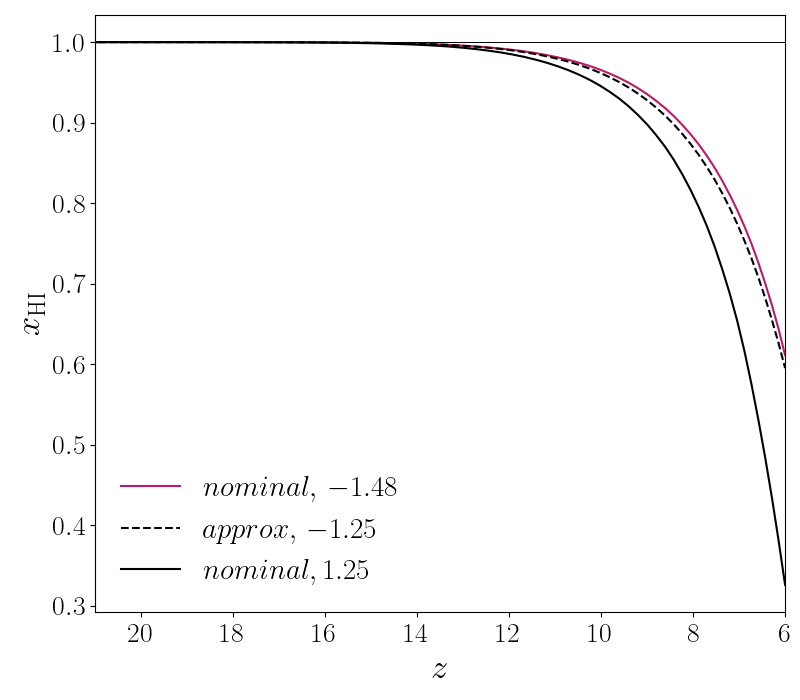}
    \caption{{\it $T_b$ and $x_{\rm HI}$ using different SFRs.} -- Global signal (left) and neutral hydrogen fraction (right). Solid lines use the {\it nominal} SFR, while dashed the {\it approximated} SFR. The magenta line shows the {\it nominal} case adopted in our analysis, where $\log_{10}$\fstar = -1.48 so that the {\it nominal} SFR agrees with the {\it approximated} one. The black, continuous line instead shows the {\it nominal} model when \fstar has the same fiducial value as MUN21. {\it The difference between the nominal and approximation models is captured by \fstar.}}
    \label{fig:global_SFR}
\end{figure*}

\begin{figure*}
    \centering
   \includegraphics[width=\columnwidth]{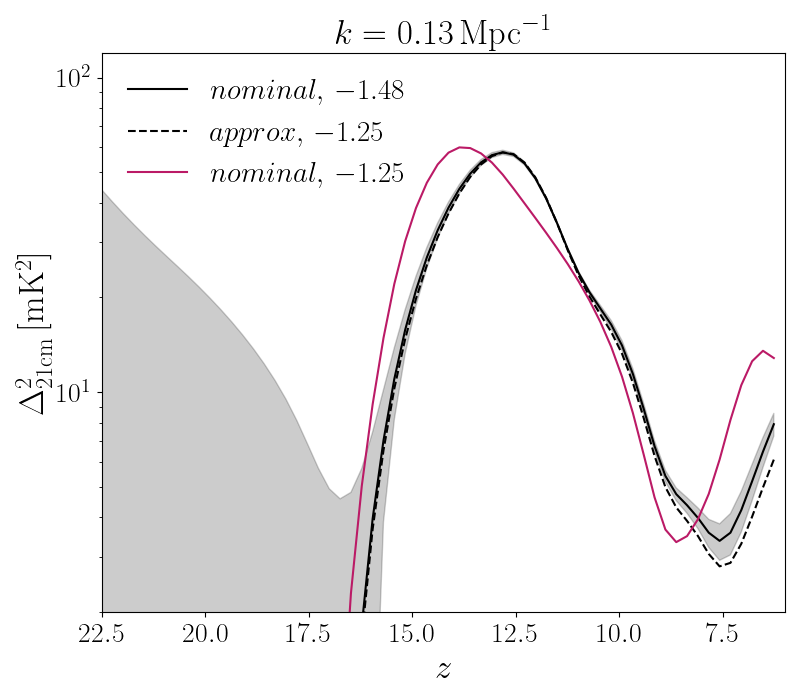}
   \includegraphics[width=\columnwidth]{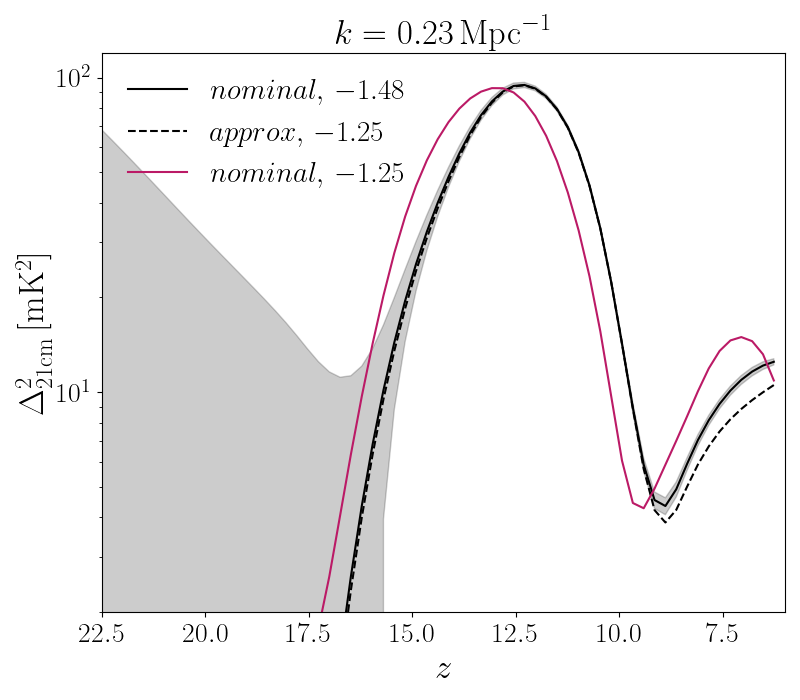}
    \caption{{\it $\Delta^2_{\rm 21cm}$ using different SFRs.} -- Power spectrum at large (left) and small (right) scales as a function of $z$. Same legend as in Fig.~\ref{fig:global_SFR}. The shaded area shows $\pm \sigma_{\rm HERA}$ with moderate foreground with respect to the {\it nominal} SFR model in the standard scenario (no FFBs). {\it The difference between the nominal and approximation models is captured by \fstar.}}
    \label{fig:pk_SFR}
\end{figure*}

\vspace*{-0.2cm}
\subsection{Effect on the 21-cm observables}

Fig.~\ref{fig:global_SFR} shows the global signal and the neutral hydrogen fraction using the {\it nominal} and {\it approximated} SFR formalism, the former using a different fiducial value for \fstar than MUN21, as described in the main text. If the same \fstar was used between the two models,
the larger SFR obtained in the {\it nominal} case would anticipate the $T_b$ peak with respect to the {\it approximated} one; consistently, reionization would be anticipated. 
However, we show that $T_b$ in the {\it nominal} case can be well recovered by simply change the normalization of the pivot value, lowering it.
Therefore, we can assume that at first order the uncertainties on the SFR model are encapsulated in the uncertainties on the parameter $f_*$. 

Analogous considerations can be made for the power spectrum, shown in Fig.~\ref{fig:pk_SFR}. In this case, changing \fstar in the {\it approximated} model we can mimic the {\it nominal} SFR power spectrum only above $z \gtrsim 10$, while at lower $z$ the {\it nominal} case anyway provides a larger power than the {\it approximated} case. We verified that such difference in the power spectrum could be reduced increasing the value of $\alpha$; however, this would also affect the global signal, increasing the difference between the {\it nominal} and {\it approximated} scenario. 
Interestingly, the differences between the two models in the power spectrum are outside the forecasted full-HERA errorbars; this implies this detector will be able to put valuable constraints on the SFR, reducing the uncertainties currently existing in the literature.

\vspace*{-0.2cm}
\subsection{FFB constraints}\label{sec:lightcones}

As a check to the reliability of our analysis, we apply the {\it approximated} SFR formalism to the study of FFB detectability. This allows a more straightforward comparison with other works in the 21-cm literature, which adopt the same prescription, e.g.,~MUN21.  
Analogously to Sec.~\ref{sec:forecasts}, we run the Fisher analysis using the {\it approximated} SFR model, with the same parameter set $\theta$. For conciseness, we only discuss $\epsilon_{\rm max} = 1$, but a similar analysis can be performed for other values.
Since the relative difference between the standard and FFB scenarios is not changed by the change in the SFR model, results using the {\it approximated} SFR depart from Sec.~\ref{sec:forecasts} $\sim\mathcal{O}(1\%)$ both under moderate and optimistic foreground assumptions. Constraints in the {\it approximated} model slightly improve since the power spectrum moves to lower $z$, where HERA is more sensitive.

\begin{figure*}
    \includegraphics[width=2\columnwidth]{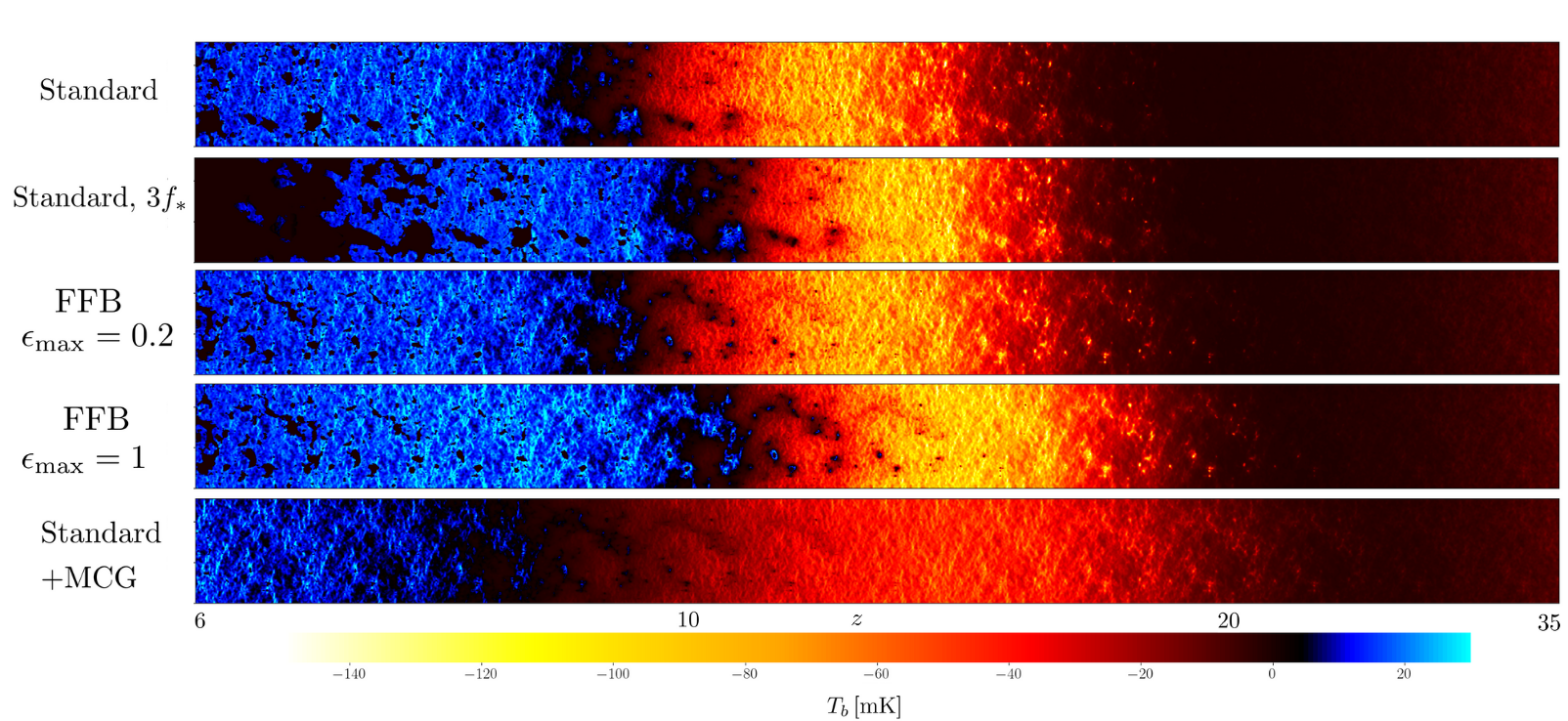}
    \caption{Lightcones produced by \texttt{21cmFAST}\,for the different models analysed in the main text (see Sec.~\ref{sec:FFB_signatures}. The colorbar indicates the value of the 21-cm global signal, while the $x$-axis shows the redshift evolution (note that smaller redshift are on the left side).}
\label{fig:lightcones}
\end{figure*}

As Fig.~\ref{fig:sfr} highlights, the difference between the {\it nominal} and {\it approximated} formalism can be reproduced in the 21-cm observables by varying the value of $f_*$; larger values of $\alpha$ could also be used to match the two models. Thus, the degeneracy between the SFR model choice and the FFB existence can be at first order understood in terms of the degeneracy between \fstar, $\alpha$ and $\epsilon_{\rm max}$. This is represented by the ellipses in Fig.~\ref{fig:ellipse} and accounted for using the marginalization in the Fisher matrix computation: given our results, HERA should be able to disentangle features in the 21-cm power spectrum due to the presence of FFBs from uncertainties in the SFR model.

\section{Lightcones}\label{app:lightcones}

In Fig.~\ref{fig:lightcones} we show the lightcone evolution of the brightness temperature in one slice of the boxes produced by \texttt{21cmFAST}. As in the main text in Sec.~\ref{sec:FFB_signatures}, we show here four different models: the ``Standard" described by Eq.~\eqref{eq:epsilon}, also in the case of large $f_*$, and the cases in which FFB are included using Eq.~\eqref{eq:eps_tot}, with $\epsilon_{\rm max} = \{0.2,1\}$. For comparison, we also show the lightcone in the standard case, when MCGs are included (see Sec.~\ref{sec:MCG}).

The behaviour of the temperature in the lightcones reflects the evolution of the global signal in Figs.~\ref{fig:GS_popII} and~\ref{fig:GS_popIII}: regions that are more yellow indicate where the peak is deeper, hence clearly the presence of FFB anticipates and deepens it. Also, it is possible to note how the presence of FFB does not strongly alter the reionization epoch. 
Finally, one can interpret the evolution of the 21-cm power spectrum described in Sec.~\ref{sec:FFB_signatures} by looking at how structures on the different scales form and evolve inside the lightcone.

\section{UV luminosity function}\label{app:UV}

Through \texttt{21cmFAST}, we estimate the luminosity function of high-$z$ galaxies in the scenarios in which FFBs are or are not included. The luminosity function $\Phi(M_{\rm UV},z)$ is defined as the number density of galaxies per UV magnitude bin, where the absolute magnitude relates to the luminosity $L_{\rm UV}$ through~\citep{Oke:1983nt}
\begin{equation}
    \log_{10}\frac{L_{\rm UV}}{{\rm erg^{-1}s^{-1}Hz^{-1}}}=0.4\times (51.63-M_{\rm UV}).    
\end{equation}
To compute it, following~\citet{Park:2018ljd} we can consider

\begin{equation}
    \Phi(M_{\rm UV},z) = f_{\rm duty}\frac{dn}{dM_h}\frac{dM_h}{dL_{\rm UV}}\frac{dL_{\rm UV}}{dM_{\rm UV}}, 
\end{equation}
where the conversion from halo mass $M_h$ to luminosity $L_{\rm UV}$ is performed knowing that
\begin{equation}
    L_{\rm UV}(M_h,z) = \frac{{\rm SFR}(M_h,z)}{1.5\times 10^{-28}M_\odot{\rm yr^{-1}}}.
\end{equation}

\begin{figure}
    \centering
    \includegraphics[width=\columnwidth]{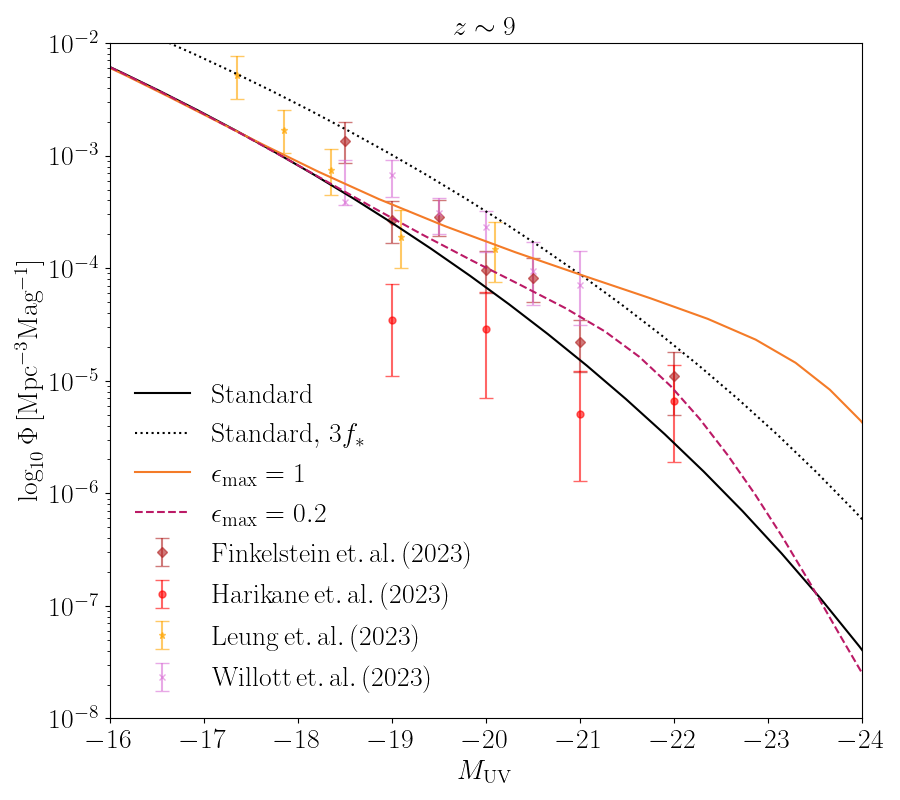}
    \caption{Luminosity function at $z\sim 9$ for the models described in the main text. The $x$-axis shows the rest frame UV absolute magnitude. We compare our results with a compilation of results from JWST~\citep{Leung_2023,harikane2023pure,finkelstein2023complete}, including lensed fields~\citep{willott2024steep}. }
    \label{fig:UV_lumfunct}
\end{figure}

The luminosity functions of the models analysed in the main text are shown in Fig.~\ref{fig:UV_lumfunct}, compared with a selection of data from HST and JWST. An extended discussion on the luminosity function in the presence of FFBs can be found in \citet{Li:2023xky}.

\end{document}